\documentclass[%
 reprint,
 amsmath,amssymb,
 aps,
]{revtex4-1}
\usepackage{graphicx}
\usepackage{dcolumn}
\usepackage{bm}
\usepackage{xcolor}
\usepackage[normalem]{ulem}
\usepackage[utf8]{inputenc}
\usepackage[T1]{fontenc}
\usepackage[section]{placeins}
\usepackage{comment}
\usepackage{mathtools}
\usepackage{subcaption}
\begin{document}
\preprint{AIP/123-QED}
	\title{Energy transfers in decaying three-dimensional Hall-magnetohydrodynamics turbulence: Dependence on
    the Ion-Inertial Length and the Magnetic Prandtl Number}
    
	\author{Pratik Patel$^1$}
       \email[]{d21ph012@phy.svnit.ac.in}
       
	\author{Sharad K. Yadav$^1$}%
	\email{sharadyadav@phy.svnit.ac.in}
	\affiliation{$^1$Department of Physics, Sardar Vallabhbhai National Institute of Technology (SVNIT), Surat 395007, Gujarat, India}

    \author{Akanksha Gupta}
    \email{akanksha@manit.ac.in}
    \affiliation{Department of Physics, Maulana Azad National Institute of Technology (MANIT), Bhopal 462003, Madhya Pradesh, India}
    
	\author{Hideaki Miura$^2$}
    \email[]{miura.hideaki@nifs.ac.jp}
	\affiliation{$^2$National Institute for Fusion Science (NIFS), Toki, Gifu 509-5292, Japan}

	\author{Rahul Pandit$^3$}
    \email[]{rahul@iisc.ac.in}
	\affiliation{$^3$Centre for Condensed Matter Theory, Department of Physics, Indian Institute of Science (IISc), Bangalore 560012, India}
	\date{\today}

    	\begin{abstract}
We present a systematic study of energy transfers in three-dimensional (3D) decaying Hall-magnetohydrodynamics (HMHD) turbulence. Our study has been designed to investigate the dependence of these transfers on initial energy spectra, the ion-inertial length $d_{i}$, and the magnetic Prandtl number $Pr_{m}$. Using shell-to-shell decomposition, we compute the energy transfer rates for (a) kinetic-to-kinetic, (b) magnetic-to-magnetic, (c) kinetic-to-magnetic, and (d) magnetic-to-kinetic interactions by carrying out extensive pseudospectral direct numerical simulations (DNSs) of 3D HMHD and its MHD counterpart. Furthermore, we examine the magnetic-to-magnetic and current-density-to-magnetic energy transfers arising from the Hall term in this framework. Energy transfers between like fields — namely, kinetic-to-kinetic and magnetic-to-magnetic — are predominantly local and forward in direction, cascading energy from large to small length scales. This behavior is consistent with a self-similar flow structure in the inertial range. Cross-field transfers, between kinetic and magnetic energy, are found to be forward but non-local in character. At low and high $Pr_{m}$, the transfers become local, and the nonlinear energy transfer is significantly suppressed. Moreover, we observe that the Hall term drives both inverse and forward transfers of magnetic energy, associated with local and non-local transfer, respectively.
	\end{abstract}
	\maketitle  

\section{Introduction}
The Hall magnetohydrodynamic (HMHD) partial differential equations (PDEs) have emerged as the hydrodynamical description of choice for a variety of conducting fluids or plasmas~\cite{lighthill1960studies,krishan1999astrophysical,miura2024formation}. The HMHD system goes beyond the PDEs of MHD or magnetohydrodynamics~\cite{1942Natur.150..405A,biskamp1997nonlinear,choudhuri1998physics,davidson2001introduction,galtier2016introduction} by introducing the Hall term, which leads to important spectral modifications at length scales smaller than the ion-inertial length $d_i$. In particular, the HMHD magnetic energy spectrum displays a Kolmogorov-type inertial range, at small wavenumbers $k$, which is like its MHD counterpart; however, for wavenumbers larger than $\simeq 1/d_i$, a second inertial range manifests itself, whose scaling exponents and their dependence on the magnetic Prandtl number have been examined recently [see, e.g., Refs.\cite{miura2019hall,miura2024formation,patel2025uncovering}]. Such studies are of relevance to many systems. Indeed, since their introduction, the HMHD equations have been used in a large variety of physical settings~\cite{Huba2003,shaikh20093d,matthaeus2012review,miura2024formation}, including magnetic-reconnection processes~\cite{yamada2010magnetic}, pulsars and neutron stars~\cite{shalybkov1997hall,hollerbach2002influence,igoshev2021evolution}, the solar wind~\cite{matthaeus1982measurement,krishan1999astrophysical,matthaeus2012review,kiyani2009global,patel2025uncovering},  the magneto-rotational instability~\cite{o2014multifluid,kopp2022influence}, sub-Alfv\'enic plasma expansion~\cite{ripin1993sub}, and magnetic-field transport in plasma opening switches~\cite{manuilenko2021magnetic}. We examine shell-to-shell energy transfers in decaying three-dimensional (3D) HMHD turbulence. We set the stage for our study by recapitulating similar work on 3D fluid and 3D MHD turbulence. Henceforth, we do not include the prefix 3D because we only study 3D systems in this paper.

Turbulence, be it hydrodynamical or MHD, is inherently nonlinear and multiscale; and it is characterized, \textit{inter alia}, by the redistribution of energy across wavenumber scales, by virtue of triadic interactions among Fourier modes of the fluid velocity $\bm u$ (or both $\bm u$ and the magnetic field $\bm b$ in MHD). In hydrodynamic turbulence, the nonlinear term of the Navier-Stokes equations couples wavenumber triads $(\bm{k}, \bm{p}, \bm{q})$, which satisfy $\bm{k}+\bm{p}+\bm{q}=\bm{0}$, and drives a forward energy cascade from large to small length scales~\cite{kraichnan1971inertial, waleffe1992nature, zhou1993interacting}. Kolmogorov's phenomenology, which treats these interactions as predominantly local in wavenumber space, yields a constant energy flux in the inertial range, and the energy spectrum $E(k) \sim k^{-5/3}$ spectrum~\cite{kraichnan1971inertial,frisch1995turbulence,verma2022variable}. In MHD turbulence, the coupling between the velocity and magnetic fields introduces additional transfer channels; shell-to-shell transfer analyses show that kinetic-to-kinetic and magnetic-to-magnetic transfers are predominantly local and forward, whereas the kinetic-to-magnetic transfers can be strongly nonlocal~\cite{verma2005energy,alexakis2005shell,debliquy2005energy}. Spectral energy transfers and flux rates are, therefore, essential for identifying the physical mechanisms behind MHD cascades~\cite{dar2001energy,verma2005energy,lessinnes2009energy, teaca2009energy,alexakis2005shell, grete2017energy, banerjee2018energy}. Not surprisingly, HMHD shell-to-shell transfers are even more complicated than their MHD counterparts. By incorporating ion-inertial scale physics, the Hall term modifies the induction equation and introduces dispersive effects that alter cascade characteristics at scales comparable to or smaller than the ion-inertial length. The analysis of triadic interactions in HMHD confirms that Hall contributions open additional transfer channels and restructure mode-to-mode interactions~\cite{banerjee2024fundamental}. In particular, the Hall term has been shown to cause back-scattering of magnetic energy~\cite{mininni2007energy, araki2011nonlocal, miura2014structure,halder2026local} toward large spatial scales, which has not been observed in MHD turbulence. 

We recall that MHD turbulence has been widely used to study many astrophysical and laboratory plasmas. Specifically, it has been used to understand the origin and amplification of magnetic fields in stars~\cite{donati2009magnetic}, galaxies~\cite{bernet2008strong}, and their clusters, and the large-scale structures of the universe~\cite{ryu2008turbulence}. At scales smaller than the ion-inertial length, the Hall term becomes significant and plays an important role in magnetic-field dynamics~\cite{huba1995hall}. Important examples include kinetic-scale turbulence and heating processes in the solar wind~\cite{krishan2004magnetic, galtier2007multiscale}, magnetic reconnection~\cite{smith2004hall}, and the solar atmosphere~\cite{krishan2004hall}. We examine the effects of the magnetic Prandtl number $Pr_m=\nu/\eta$ on the energy transfer mechanism in 3D Hall-MHD turbulence, where $\nu$ and $\eta$ are the kinematic viscosity and magnetic diffusivity, respectively. The magnetic Prandtl number controls the ratio of viscous and resistive dissipative length scales. In some astrophysical systems, the magnetic Prandtl number is very low; e.g., it is $\sim 10^{-7}-10^{-4}$ in the sun's convection zone~\cite{brandenburg2005astrophysical} and $\sim 10^{-5}$ in the liquid-metal cores of planets~\cite{ponty2005numerical}. In laboratory systems, $Pr_m$ is  $10^{-5}$ for liquid-metal dynamo experiments~\cite{gailitis2002colloquium}, whereas it is $\sim10^{-3}-5\times 10^{2}$ for plasma Couette-flow experiments~\cite{ebrahimi2011global}. In contrast, some astrophysical plasmas exhibit turbulent flow at very high magnetic Prandtl numbers. $Pr_m$ is estimated to be in the range of $\sim10^{4}-10^{20}$ for the interstellar mediums (ISM)~\cite{ponty2005numerical,ferriere2020plasma}, $\sim10^{10}$  for molecular clouds~\cite{schekochihin2007fluctuation}, and $\sim 10^{29}$ for the intracluster medium (ICM)~\cite{schekochihin2005plasma}. $Pr_m$ is nearly one for slow solar-wind turbulence at 1 au~\cite{hand2026empirical}. \\
At scales smaller than the ion-inertial length (i.e., lengths $<d_i$), the scaling exponents of the kinetic- and magnetic-energy spectra exhibit magnetic-Prandtl-number dependence and a correlation with each other~\cite{miura2019hall, patel2025uncovering}.  Shell-to-shell energy transfers in HMHD turbulence have not been characterized systematically, hitherto, for different values of $Pr_m$. Note that the value of $Pr_m$ can influence dynamo action~\cite{brandenburg2009large, sahoo2010dynamo,malyshkin2010magnetic,brandenburg2019reversed}, cascade properties, and scale interactions in MHD~\cite{alexakis2005shell,mininni2005shell, sahoo2011systematics}.

High-resolution direct numerical simulations (DNSs) of 3D HMHD turbulence, e.g., with $1024^3$ collocation points, become computationally challenging when  $Pr_m$ differs significantly from unity, because the velocity and magnetic-field dissipative ranges are widely separated. We present a systematic investigation of shell-to-shell energy transfer in 3D HMHD turbulence for $Pr_m=0.1$, $1$, and $10$. Using the mode-to-mode transfer formalism~\cite{mkverma2019energytransfer}, we quantify inter-shell transfers within and between the velocity and magnetic fields.
\par
Before we present the details of our work, we summarize our principal results qualitatively. The energy transfers because of the nonlinear terms, $(\bm{u}\cdot \nabla)\bm{u}$ and $ (\bm{u}\cdot \nabla)\bm{b}$ [Eqs.~\eqref{eq:vel} and ~\eqref{eq:mag}] exhibit dominant local forward cascades in 3D HMHD turbulence, at all length scales irrespective of the values of the magnetic Prandtl number $Pr_{m}$ and the ion-length scales $d_{i}$ (including MHD case) considered in our DNSs; by contrast, the cross-field energy-transfer terms, $(\bm{b} \cdot \nabla)\bm{b}$ and $ (\bm{b} \cdot \nabla)\bm{u}$ [Eqs.~\eqref{eq:vel} and ~\eqref{eq:mag}] shows signatures of non-local transfer if $Pr_m = 1$. Furthermore, for $Pr_{m} = 0.1$ and $10$, these non-local transfers are substantially suppressed and are observable only over narrow inertial ranges. The energy transfer, which is driven by the Hall term [$d_{i}(\bm{j} \cdot \nabla)\bm{b}$ - $d_{i}(\bm{b} \cdot \nabla)\bm{j}$ in Eq.~\eqref{eq:mag}], exhibits a considerable local inverse transfer of magnetic energy at large spatial scales. Thus, our study brings out qualitative changes in energy transfers; we quantify the dependence of the changes on $Pr_m$ and $d_i$ in 3D HMHD .
\par
The remainder of this paper is organized as follows: In Section~\ref{sec:methods}, we describe the numerical equation that we use to examine energy transfers in Fourier space, mode-to-mode energy transfers, and shell-to-shell energy transfers; we also give the numerical methods we use and the parameters for our simulations. Section~\ref{sec:result}, is devoted to a presentation of our results. Section \ref{sec:conclusion} contains a discussion of the significance of our results.

\section{Governing equations and Numerical Methods}\label{sec:methods}
We work with the incompressible three-dimensional (3D) Hall magnetohydrodynamic (HMHD) equations~\eqref{eq:vel}-\eqref{eq:divmag} given below:
	\begin{gather}
		\label{eq:vel}
		\frac{\partial\bm{u}}{\partial\textit{t}}=-\nabla\bar{P}-(\bm{u}\cdot\nabla)\bm{u}+ (\bm{b}\cdot\nabla)\bm{b}+\nu\nabla^2\bm{u}+\bm{f}_u\,;\\
        \begin{aligned}
        \label{eq:mag} 
        \frac{\partial \bm{b}}{\partial t}
        ={}&-(\bm{u}\cdot\nabla)\bm{b}
        +(\bm{b}\cdot\nabla)\bm{u}+d_i(\bm{j}\cdot\nabla)\bm{b}\\
        &-d_i(\bm{b}\cdot\nabla)\bm{j}
        +\eta\nabla^2\bm{b}
        +\bm{f}_b\,;
        \end{aligned}
        \\
		\label{eq:divvel}
		\nabla\cdot\bm{u}=0\,;\\
		\label{eq:divmag}
		\nabla\cdot\bm{b}=0\,.
	\end{gather}
 Equations~\eqref{eq:vel}-\eqref{eq:divmag} govern the spatiotemporal evolution of the velocity field $\bm{u}$ and the magnetic field $\bm{b}$~\cite{krishan1999astrophysical,miura2019hall,yadav2022statistical}; here, $\bar{P}$
denotes the total pressure $(p+b^2/8\pi)$, $\bm{j}=\nabla\times\bm{b}$ is the current density, $d_i$
is the ion-inertial length, which appears as the coefficient of the Hall term $(\bm{j}\times \bm{b})$ [for  $d_{i}=0$, the HMHD model~\eqref{eq:vel}-\eqref{eq:divmag}
reduces to incompressible MHD]; $\nu$ and $\eta$ are the kinematic viscosity and magnetic diffusivity, respectively. Equations~\eqref{eq:divvel} and \eqref{eq:divmag} enforce incompressibility and the solenoidal constraint on the magnetic field. The external forces $\bm{f}_u$ and $\bm{f}_b$ are set to zero if we consider decaying 3D HMHD turbulence. 
\subsection{Mode-to-Mode Energy Transfers via Triad Interactions}
\label{subsec:mode2mode}
If we write Eqs.~\eqref{eq:vel}-\eqref{eq:divmag} in Fourier space (see below), we find that different Fourier modes interact and exchange energy through the nonlinear terms in the governing equations; informally, this is referred to as the interaction of eddies of different sizes. As we show presently, these interactions involve wave-number triads $(\bm{k}, \bm{p}, \bm{q})$ and the full nonlinear energy transfer is the sum over all triad combinations that satisfy the constraint $\bm{k}+\bm{p}+\bm{q}=\bm{0}$. Traditionally, the combined energy transfer to a mode from the other two members of a triad has been treated as the fundamental quantity. Batchelor~\cite{batchelor1950spontaneous} first proposed a formula for energy transfer between regions of Fourier space in fluid turbulence, and Domaradzki and Rogallo~\cite{domaradzki1990local} used this framework to compute shell-to-shell transfer rates. Dar, \textit{et al.}~\cite{dar2001energy} used a different approach for the mode-to-mode transfer formalism and computed shell-to-shell transfers and spectral fluxes.\\
We adapt the formalism of Ref.~\cite{dar2001energy} for our study of the 3D HMHD system~\eqref{eq:vel}-\eqref{eq:divmag}. We express any vector field $\bm{a}(\bm{x},t)$ in the Fourier representations 
\begin{equation}
\bm{a}(\bm{x},t)=\sum_{\bm{k}} \hat{\bm a}(\bm{k},t)\exp(i\bm{k}\cdot\bm{x})\,,
\end{equation}
where $\hat{\bm{a}}(\bm{k},t)$ is the Fourier amplitude of the given field; $\bm{a}(\bm{x},t)$ can be either $\bm{u}$ or $\bm{b}$. For notational convenience, we suppress the caret with the understanding that the wave-vector argument implies that we are considering a spatial Fourier transform.  
In Fourier space the 3D HMHD system~\eqref{eq:vel}-\eqref{eq:divmag} is:
\begin{equation}
\begin{aligned}
\frac{ d\bm{u}(\bm{k})}{dt}
={}&-i\bm{k}\,p(\bm{k})-i\sum_{\bm{p}}
\left[\bm{k}\cdot\bm{u}(\bm{k}-\bm{p})\right]
\bm{u}(\bm{p})
\\
&+i\sum_{\bm{p}}
\left[\bm{k}\cdot\bm{b}(\bm{k}-\bm{p})\right]
\bm{b}(\bm{p})-\nu k^2\bm{u}(\bm{k})\,;
\end{aligned}
\label{Eq. 1.20}
\end{equation}
\begin{equation}
\begin{aligned}
\frac{d\bm{b}(\bm{k})}{dt}
={}&-i\sum_{\bm{p}}
\left[\bm{k}\cdot\bm{u}(\bm{k}-\bm{p})\right]
\bm{b}(\bm{p})
\\
&+i\sum_{\bm{p}}
\left[\bm{k}\cdot\bm{b}(\bm{k}-\bm{p})\right]
\bm{u}(\bm{p})
\\
&+id_i\sum_{\bm{p}}
\left[\bm{k}\cdot\bm{j}(\bm{k}-\bm{p})\right]
\bm{b}(\bm{p})
\\
&-id_i\sum_{\bm{p}}
\left[\bm{k}\cdot\bm{b}(\bm{k}-\bm{p})\right]
\bm{j}(\bm{p})
\\
&-\eta k^2\bm{b}(\bm{k})\,;
\end{aligned}
\label{Eq. 1.21}
\end{equation}
\vspace{-0.3cm}
\begin{equation}
i\bm{k}\cdot\bm{u}(\bm{k})=0\,;
\label{1.22}
\end{equation}
\vspace{-0.6cm}
\begin{equation}
i\bm{k}\cdot\bm{b}(\bm{k})=0\,.
\label{1.23}
\end{equation}
To derive the kinetic and magnetic energy transfer rates, we multiply Eq.~\eqref{Eq. 1.20} by $\bm{u}^*(\bm{k})$ and Eq.~\eqref{Eq. 1.21} by $\bm{b}^*(\bm{k})$, then add each result to its complex conjugate; after simplification we obtain:

\begin{equation}
\begin{aligned}
\frac{dE_u(\bm{k})}{dt}
={}&
\underbracket{
-\Re\!\left(
i[\bm{k}\!\cdot\!\bm{u}(\bm{q})]
[\bm{u}(\bm{p})\!\cdot\!\bm{u}^*(\bm{k})]
\right)
}_{\displaystyle S_{uu}(\bm{k}|\bm{p}|\bm{q})}
\\[5pt]
&+
\underbracket{
\Re\!\left(
i[\bm{k}\!\cdot\!\bm{b}(\bm{q})]
[\bm{b}(\bm{p})\!\cdot\!\bm{u}^*(\bm{k})]
\right)
}_{\displaystyle S_{bu}(\bm{k}|\bm{p}|\bm{q})}
\\[5pt]
&-2\nu k^2E_u(\bm{k})\,;
\end{aligned}
\end{equation}

\begin{equation}
\begin{aligned}
\frac{dE_b(\bm{k})}{dt}
={}&
\underbracket{
-\Re\!\left(
i[\bm{k}\!\cdot\!\bm{u}(\bm{q})]
[\bm{b}(\bm{p})\!\cdot\!\bm{b}^*(\bm{k})]
\right)
}_{\displaystyle S_{bb}(\bm{k}|\bm{p}|\bm{q})}
\\[5pt]
&+
\underbracket{
\Re\!\left(
i[\bm{k}\!\cdot\!\bm{b}(\bm{q})]
[\bm{u}(\bm{p})\!\cdot\!\bm{b}^*(\bm{k})]
\right)
}_{\displaystyle S_{ub}(\bm{k}|\bm{p}|\bm{q})}
\\[5pt]
&+
\underbracket{
\Re\!\left(
i[\bm{k}\!\cdot\!\bm{j}(\bm{q})]
[\bm{b}(\bm{p})\!\cdot\!\bm{b}^*(\bm{k})]
\right)
}_{\displaystyle S^{\mathrm{Hall}}_{bb}(\bm{k}|\bm{p}|\bm{q})}
\\[5pt]
&-
\underbracket{
\Re\!\left(
i[\bm{k}\!\cdot\!\bm{b}(\bm{q})]
[\bm{j}(\bm{p})\!\cdot\!\bm{b}^*(\bm{k})]
\right)
}_{\displaystyle S^{\mathrm{Hall}}_{jb}(\bm{k}|\bm{p}|\bm{q})}
\\[5pt]
&-2\eta k^2 E_b(\bm{k})\,;
\end{aligned}
\end{equation}
here, $E_u(\bm{k})=|\bm{u}(\bm{k})|^2/2$ and $E_b(\bm{k})=|\bm{b}(\bm{k})|^2/2$ are, respectively, the kinetic and magnetic energy spectra. The total-pressure term vanishes by virtue of the incompressibility and solenoidal constraints~\eqref{1.22}-\eqref{1.23}). Terms of the type $S_{XY}$ denote the transfer of energy from the field $X$, mode $\bm{p}$, to the field $Y$, mode $\bm{k}$, via the mediator mode $\bm{q}$ [see Ref.~\cite{stanisic2012mathematical}]. These nonlinear terms obey the following conditions~\cite{kraichnan1959structure, lesieur1987turbulence}:
\begin{equation}
S_{XY}(\bm{k}|\bm{p}|\bm{q})+	S_{XY}(\bm{p}|\bm{k}|\bm{q})+S_{XY}(\bm{q}|\bm{k}|\bm{p})=0\,,
\end{equation}
where $X$ and $Y$ can be either $\bm{u}$ or $\bm{b}$.

\subsection{Flux Rates and Shell-to-Shell Energy Transfers}
\label{subsec:fluxes}
Given $S_{XY}(\bm{k}|\bm{p}|\bm{q})$, we can calculate the flux 
\begin{equation}
    \Pi^{X^{<}}_{Y^{>}}(k_0)=\sum_{p\le k_0}\sum_{k > k_0}S_{XY}(\bm{k}|\bm{p}|\bm{q})\,,
\end{equation}
which is important in the inertial range.
This flux is the energy transfer from inside of the wavenumber sphere, of radius $k_0$, for the field $X$, to the region outside the same sphere, but for the field $Y$. We focus mainly on the following three fluxes:

\begin{equation}
\begin{aligned}
\Pi_u(\bm{k}_0)
={}&
\underbracket{\sum_{p\le k_0}\sum_{k > k_0}S_{uu}(\bm{k}|\bm{p}|\bm{q})
}_{\displaystyle \Pi^{u<}_{u>}(\bm{k}_0)}
\\[5pt]
&+
\underbracket{\sum_{p\le k_0}\sum_{k > k_0}S_{bu}(\bm{k}|\bm{p}|\bm{q})
}_{\displaystyle \Pi^{b<}_{u>}(\bm{k}_0)}\,;
\end{aligned}
\label{Eq:14}
\end{equation}

\begin{equation}
\begin{aligned}
\Pi_b(\bm{k}_0)
={}&
\underbracket{\sum_{p\le k_0}\sum_{k > k_0}S_{bb}(\bm{k}|\bm{p}|\bm{q})
}_{\displaystyle \Pi^{b<}_{b>}(\bm{k}_0)}
\\[5pt]
&+
\underbracket{\sum_{p\le k_0}\sum_{k > k_0}S_{ub}(\bm{k}|\bm{p}|\bm{q})
}_{\displaystyle \Pi^{u<}_{b>}(\bm{k}_0)}\,;
\end{aligned}
\label{Eq:15}
\end{equation}

\begin{equation}
\begin{aligned}
\Pi_H(\bm{k}_0)
={}&
\underbracket{d_i\sum_{p\le k_0}\sum_{k > k_0}S^{\mathrm{Hall}}_{bb}(\bm{k}|\bm{p}|\bm{q})
}_{\displaystyle \Pi^{b<}_{b>}(\bm{k}_0)(\mathrm{Hall})}
\\[5pt]
&+
\underbracket{d_i\sum_{p\le k_0}\sum_{k > k_0}S^{\mathrm{Hall}}_{jb}(\bm{k}|\bm{p}|\bm{q})
}_{\displaystyle \Pi^{j<}_{b>}(\bm{k}_0)(\mathrm{Hall})}\,.
\end{aligned}
\label{Eq:16}
\end{equation}

The sum of these fluxes is
\begin{equation}
    \Pi_{T}(k_0)=\Pi_u(k_0)+\Pi_b(k_0)+\Pi_{H}(k_0)\,.\label{Eq:17}
\end{equation}
If we sum over all possible triadic interactions, we obtain the shell-to-shell transfer rate~\cite{dar2001energy,batchelor1950spontaneous}:
\begin{equation}
	T^{mn}_{uu}=\frac{1}{2}\sum_{\bm{k}\in n}\sum_{\bm{p}\in m}S_{uu}(\bm{k}|\bm{p}|\bm{q})\,,
\end{equation}
where the sum $\bm{k}$ runs over the shell $n$, the sum $\bm{p}$ runs over the shell $m$, and $\bm{q}=\bm{k}-\bm{p}$. The different shell-to-shell energy transfer rates from shell $m$ to shell $n$ are defined below:
\begin{gather}
	T^{mn}_{uu}=\sum_{k_n<k<k_{n+1}}\sum_{k_m<p<k_{m+1}}S_{uu}(\bm{k}|\bm{p}|\bm{q})\,;\\
 	T^{mn}_{bb}=\sum_{k_n<k<k_{n+1}}\sum_{k_m<p<k_{m+1}}S_{bb}(\bm{k}|\bm{p}|\bm{q})\,;\\
	T^{mn}_{ub}=\sum_{k_n<k<k_{n+1}}\sum_{k_m<p<k_{m+1}}S_{ub}(\bm{k}|\bm{p}|\bm{q})\,;\\
	T^{mn}_{bu}=\sum_{k_n<k<k_{n+1}}\sum_{k_m<p<k_{m+1}}S_{bu}(\bm{k}|\bm{p}|\bm{q})\,;\\
    H^{mn}_{bb}=d_i\sum_{k_n<k<k_{n+1}}\sum_{k_m<p<k_{m+1}} S^{Hall}_{bb}(\bm{k}|\bm{p}|\bm{q})\,; \label{Eq:23}\\
H^{mn}_{jb}=d_i\sum_{k_n<k<k_{n+1}}\sum_{k_m<p<k_{m+1}}S^{Hall}_{jb}(\bm{k}|\bm{p}|\bm{q})\,.\label{Eq:24}   
\end{gather}
Shell-to-shell energy transfers between velocity and magnetic variables are well-defined; but separating forward from backward contributions is not straightforward. The net transfer between two shells is the sum of many triad interactions, some of which increase the energy of the shell $n$ and others that decrease it. The shell-to-shell transfer is, therefore, a sum of positive $T^{mn,(+)}_{ub}>0$ and negative $T^{mn,(-)}_{ub}<0$ contributions. When positive contributions dominate ($T^{mn}_{ub}=T^{mn,(+)}_{ub}+T^{mn,(-)}_{ub}>0$) and $n>m$, the net interaction yields a direct energy cascade. If $T^{mn}_{ub}<0$ for $n>m$, we have an inverse cascade of energy.

\subsection{Direct Numerical Simulations (DNSs):}
\label{subsec:dns}

\begin{figure}[!h]
	\centering
    \includegraphics[scale=0.30]{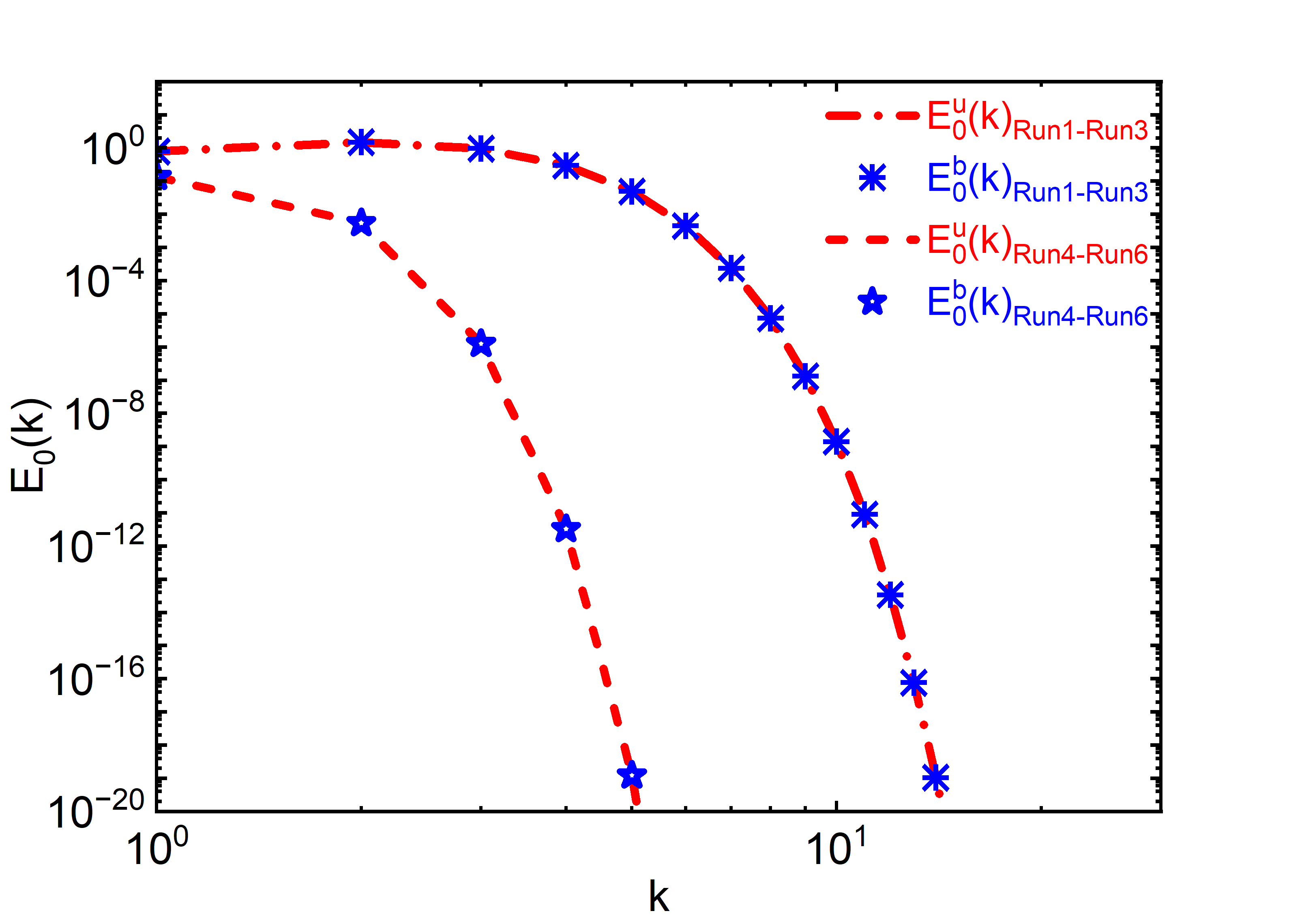}
	\caption{\small Log-log plots, versus the wavenumber $k$, of initial spectra for the kinetic (red) and magnetic (blue) energies for Run1-Run6.} 
    \label{fig:initialenrgyspectra}
\end{figure}

We solve Eqs.~\eqref{eq:vel} and ~\eqref{eq:mag} numerically using a pseudospectral direct numerical simulation (DNS) in a cubical box of side $2\pi$, with periodic boundary conditions along all three spatial directions. The Fourier transform of a nonlinear term involves a convolution, which is computationally expensive, so in this DNS, we use the pseudospectral method~\cite{canuto2012spectral} to compute the nonlinear terms efficiently. The aliasing errors that occur during the evaluation of nonlinear terms are removed using the $2/3$ de-aliasing rule. In our DNSs, the number of collocation points is $N^3=1024^3$; so the maximal wavenumber is $k_{max}\simeq 341$. For temporal advancement of the field variables, we employ the second-order, slaved Adams-Bashforth method. We have carried out 6 sets of DNSs, each one of them with $1024^3$ collocation points. The parameters used in our DNSs are listed in Table~\ref{table1}. We give below an overview of our DNSs:
\begin{itemize}
    \item 
    In Run1, Run2, and Run3, the initial energy spectra are $E_{0}^{u,b}\sim k^{2} \exp(-k^{2}/k_{0}^2)$ ($k_{0}=2$); in Run4, Run5, and Run6, we use $E_{0}^{u,b} \sim k^{4}\exp(-2k^{2})$ [see Fig.~\ref{fig:initialenrgyspectra}].
    \item 
    The first three runs [Run1-Run3] are carried out at $Pr_{m}=1.0$, but with different ion-inertial length\\

\begin{table*}[ht]
	\begin{center}
		\setlength{\tabcolsep}{13pt}		
		\begin{tabular}{c c c c c c c c c c}
			\hline
			\hline
Runs & N & $\nu$ & $Pr_m$ & $d_i$ & $\delta t$ & $Re_{\lambda}$  & $\tau_c$ & $k_{max}\eta^u_d$ & $k_{max}\eta^u_d$ \\
							\hline

                            Run1 & 1024 & $5\times10^{-4}$ & 1.0 & 0.00 & $10^{-4}$ & 143 & 0.5 & 1.53 & 1.37\\

                            Run2 & 1024 & $5\times10^{-4}$ & 1.0 & 0.025 & $10^{-4}$ & 153 & 0.5 & 1.60 & 1.30\\ 

                            Run3 & 1024 & $5\times10^{-4}$ & 1.0 & 0.05 & $10^{-4}$ & 176 & 0.5 & 1.71 & 1.25\\
                            
							Run4 & 1024 & $5\times10^{-4}$ & 0.1 & 0.05 & $5\times10^{-5}$ & 211.38 & 6.93 & 3.83 & 13.72\\
							
							Run5 & 1024 &  $5\times10^{-4}$ & 1 & 0.05 & $10^{-5}$ & 129.83 & 8.23 & 3.16 & 2.72\\
							
							Run6 & 1024 &  $5\times10^{-3}$ & 10 & 0.05 & $10^{-5}$ & 28.30 & 7.21 & 15.40 & 3.16\\

			\hline
			\hline
		\end{tabular} 
	\end{center}
	\caption{\small List of parameters used in our DNS runs, [Run1-Run6];
    $N$,$\nu$, $Pr_{m}$, $d_{i}$, $\delta t$ are, respectively, number of collocation points, kinematic viscosity, magnetic Prandtl number, ion-length scale, time-step; $Re_{\lambda}$, $\eta_{d}^{u}$ and $\eta_{d}^{b}$ are kinetic Reynolds number, kinetic dissipation length scale and magnetic dissipation length scale, respectively, computed at cascade completion time $\tau_{c}$. Simulations correspond to the following physical systems, Run1 $\longrightarrow$ MHD
    and
    [Run2-Run6] $\longrightarrow$ HMHD.
    }
	\label{table1}
\end{table*}

\noindent
scales, namely, for $d_{i}=0.0$ [Run1], $0.025$ [Run2], and $0.05$ [Run3].  By contrast, in Run4-Run6, we fix $d_{i}=0.05$, but we vary the magnetic Prandtl number:  $Pr_{m}=0.1$ [Run4], $1.0$ [Run5], and $10.0$ [Run6].    
    \item 
    The values of $k_{max}\eta_{d}^{u}$ and $k_{max}\eta_{d}^{b}$, calculated at the cascade-completion time $\tau_{c}$, are always greater than unity; this indicates that all important length scales present are well resolved; $k_{max}=N/3\simeq 341.3$ and $\eta_{d}^{u(b)}$ is the kinetic (magnetic) dissipation length scale. 
\end{itemize}

Data from our DNS runs  [Run1-Run6] have already been analyzed and published, but in different contexts [see, e.g., Refs.\cite{miura2014structure, yadav2022statistical,PhysRevE.100.063207,miura2024formation,patel2025uncovering}], where we observed two scaling ranges. In the first, energy spectra displayed Kolmogorov-type (K41) scaling, i.e., $E_u(k)\sim k^{-5/3}$ and $E_b(k)\sim k^{-5/3}$, in the inertial region $k \ll k_{i}$, where $k_{i} \sim 1/d_{i}$ for all the values of $Pr_m$ we considered. In the second-inertial region, $k_{i}<k \ll k_d$, we showed that the spectra $E_u(k)$ and $E_b(k)$ exhibit power-law scaling with exponents that depend on $Pr_m$; the magnetic energy dominates over the kinetic energy for $Pr_m\ge 1$, but the kinetic energy increases significantly for $Pr_m <  1 $. Moreover, we explored signatures of intermittency~\cite{yadav2022statistical,patel2025uncovering} and the formation of structures~\cite{miura2024formation} in 3D HMHD turbulence.

\section{Results}\label{sec:result}

We now present the results of our DNS investigation of the dependence on the magnetic Prandtl number $Pr_m$ and the ion-inertial length $d_i$ of shell-to-shell energy transfer in 3D HMHD turbulence. In particular, we quantify how the different nonlinear terms in the incompressible 3D HMHD equations~\eqref{eq:vel}-\eqref{eq:divmag} contribute to scale-by-scale energy transfer between the velocity and magnetic fields. We concentrate on decaying turbulence (see Table~\ref{table1} and Refs.~\cite{yadav2022statistical,patel2025uncovering}). In Section~\ref{subsec:fluxrates} we examine flux rates; then we turn to our DNS results for shell-to-shell energy transfers in Section~\ref{subsec:shell-shell}.

\subsection{Flux Rates}
\label{subsec:fluxrates}

\begin{figure*}[!htbp]
	\centering
     \includegraphics[width=1.6\columnwidth]{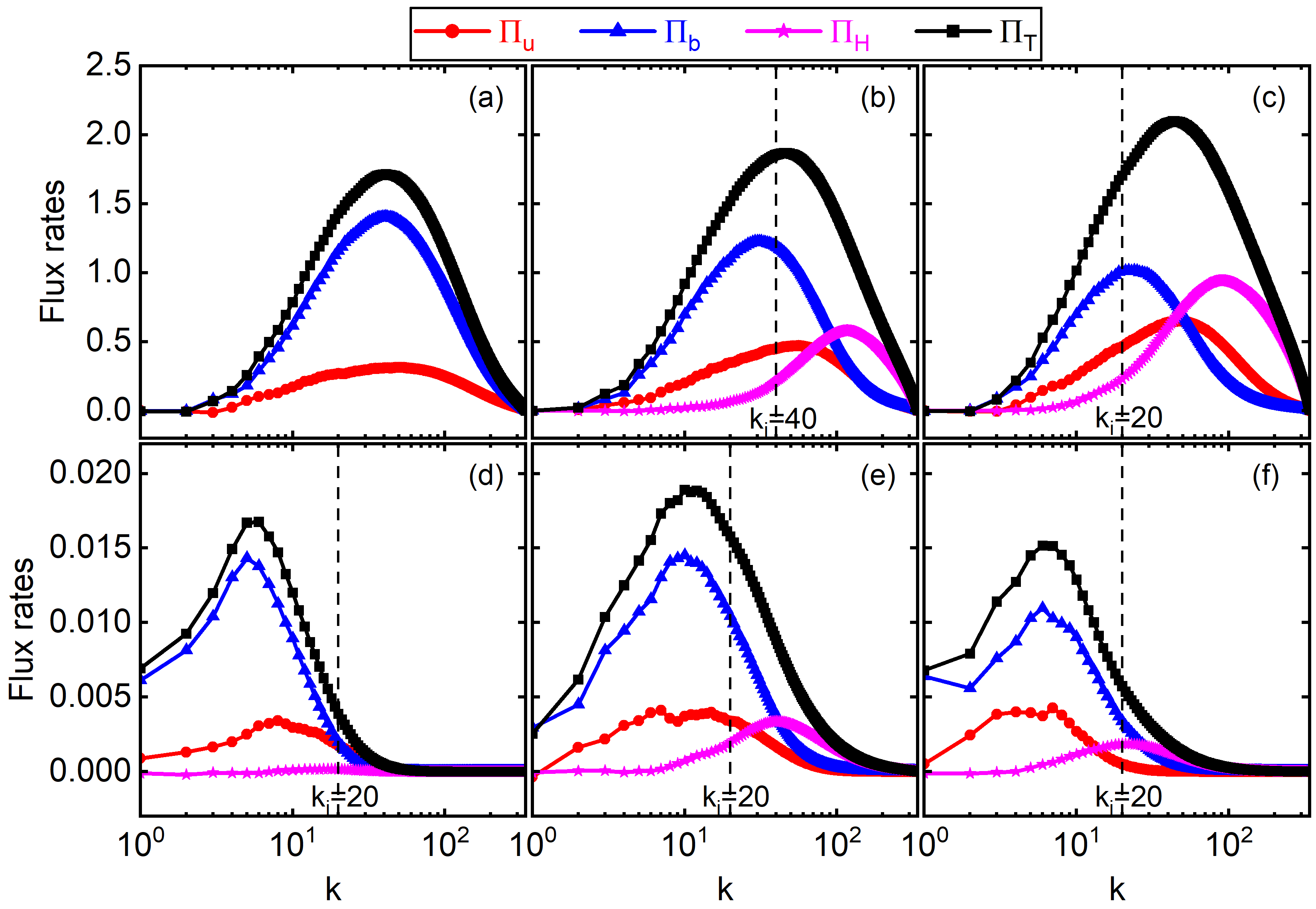}
	\caption{\small Plots of different energy fluxes rates $\Pi_u$ (Red curve), $\Pi_b$ (Blue curve), $\Pi_{H}$ (Magenta curve) and $\Pi_{T}$ (Black curve) for (a) Run1, (b) Run2, (c) Run3, (d) Run4, (e) Run5, and (f) Run6}
    \label{fig:flux_rates_Run123456}
\end{figure*}

\begin{figure*}[!htbp]
	\centering
    \includegraphics[width=1.6\columnwidth]{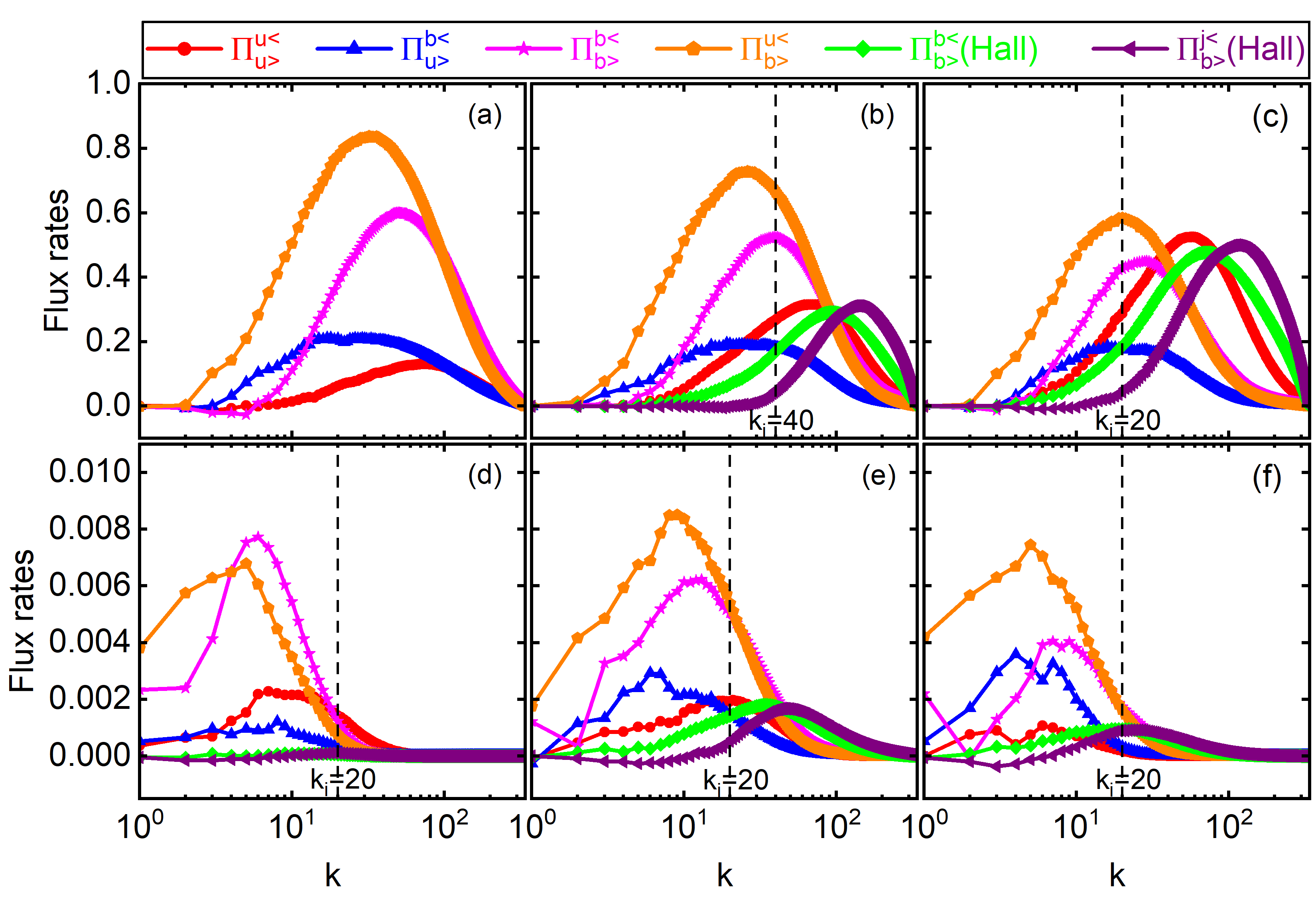}
	\caption{\small Plots of different energy flux rates $\Pi_{u>}^{u<}$ (Red curve), $\Pi_{u>}^{b<}$ (Blue curve), $\Pi_{b>}^{b<}$ (Magenta curve),  $\Pi_{b>}^{u<}$ (Orange curve), $\Pi_{b>}^{b<}{(Hall)}$ (Green curve) and $\Pi_{b>}^{j<}(Hall)$ (Purple curve) for (a) Run1, (b) Run2, (c) Run3, (d) Run4, (e) Run5, and (f) Run6}
    \label{fig:termwise_flux_rates_Run123456}
\end{figure*}

We compute the energy fluxes from our DNSs, using Eqs.~\eqref{Eq:14}-\eqref{Eq:17}, and present semi-log plots of $\Pi_u$ (red), $\Pi_b$ (blue), $\Pi_{H}$ (magenta), and $\Pi_{T}$ (black) versus the wavenumber $k$ in Figures \ref{fig:flux_rates_Run123456} (a)-(f) for Run1-Run6, respectively. By comparing Figures \ref{fig:flux_rates_Run123456} (a), (b), and (c) for Run1, Run2, and Run3, respectively, we see how these fluxes depend upon the ion-inertial scale $d_{i}$. Similarly, we can contrast Figures \ref{fig:flux_rates_Run123456} (d), (e), and (f) for Run4, Run5, and Run6, respectively, to see how these fluxes change with the magnetic Prandtl number $Pr_m$. For a comparison between 3D HMHD and 3D MHD, recall that Run1 is the control 3D MHD simulation, because $d_{i}=0$.  

Observe that $\Pi_u$, $\Pi_b$, $\Pi_{H}$, and $\Pi_{T}$ are positive for all wavenumbers up until the dissipation wavenumber, which indicates that energy is continuously leaving from modes, inside the sphere of radius $k_0$, to wavenumbers $ > k_0$ (i.e., we have a forward cascade of energy). $\Pi_u$ and $\Pi_b$ start to increase from small wavenumbers and then display peaks at intermediate wavenumbers. In Run1, the peak appears in the inertial region, whereas in Run2 and Run3, the peak appears at a wavenumber close to or beyond $k_{i}$. By contrast, in Run4-Run6, the peaks in $\Pi_{u}$ and $\Pi_{b}$ in the inertial region before $k_{i}$. Moreover, $\Pi_{b}$ dominates over $\Pi_{u}$ in all our runs and for all wavenumbers. Note that in Run1-Run3, $\Pi_{u}$ increases as we enhance $d_{i}$ and thus the strength of the Hall; however, $\Pi_{b}$ shows the opposite trend, i.e., it decreases with increasing $d_{i}$. 

Next we explore the energy flux $\Pi_{Hall}$ (purple) because of the Hall term: For wavenumbers $k \ll k_{i}$, the value of $\Pi_{Hall}$ is quite small (almost zero) but it starts increasing, as $k$ approaches $k_{i}$, and it then has a peak in the second inertial region, which depends upon $d_{i}$ and $Pr_{m}$, because $[k_{i} \sim 1/d_i$ and $k^{\eta_{d}^{b}}$, the wavenumber corresponding to the magnetic dissipation length scale ${\eta_{d}^{b}}$, is proportional to $Pr_{m}$. In our DNSs of 3D HMHD, the second inertial ranges $[k_{i} - k^{\eta_{d}^{b}}]$ are as follows: Run2:$[40 - 262.30]$, Run3:$[20 - 272.80]$, Run4:$[20 - 24.93]$, Run5:$[20 - 126.58]$, and Run6:$[20 - 111.10]$. Note that this range is very narrow in Run4; it hardly spans $\simeq 4$ wavenumbers, i.e., the second inertial region is suppressed by the dissipation range, so there is no clear peak in $\Pi_{Hall}$ for Run4.
\par 
Furthermore, we have investigated the contribution of individual terms that appear in the energy fluxes, $\Pi_{u}$, $\Pi_{b}$, and $\Pi_{Hall}$ [see Eqs.~(16)-(18)]; their wavenumber-dependences are displayed in Fig.~\ref{fig:termwise_flux_rates_Run123456} for all our runs.
Note that $\Pi_{u}$ is the sum of  
$\Pi^{u^{<}}_{u^{>}}$ and $\Pi^{b^{<}}_{u^{>}}$, where the former corresponds to the advection term 
$\sim (\bm{u} \cdot {\nabla}) \bm{u}$ and 
latter to the Lorentz term $\sim (\bm{b} \cdot 
{\nabla}) \bm{u}$ that appear in Eq.~\ref{eq:vel};  similarly, the 
energy flux $\Pi_{b} $ is the sum of $\Pi^{b^{<}}
_{b^{>}}$ and $\Pi^{u^{<}}_{b^{>}}$ in which $
\Pi^{b^{<}}_{b^{>}}$ stands for the term $\sim 
(\bm{u} \cdot \nabla) \bm{b}$ and $\Pi^{u^{<}}
_{b^{>}}$ for $\sim (\bm{b} \cdot \nabla) 
\bm{u}$ of Eq.~\ref{eq:mag}; and $\Pi_{Hall}$ 
is the sum of $\Pi^{b^{<}}_{b^{>}} $ and $\Pi^{j^{<}}
_{b^{>}}$, where the former and the latter correspond, respectively, to the terms $\sim d_{i} 
(\bm{j} \cdot \bm{\nabla}) \bm{b}$ and $\sim d_{i} (\bm{b} \cdot \nabla) 
\bm{J}$ in Eq.~\ref{eq:mag}. The top panel of 
Fig.~\ref{fig:termwise_flux_rates_Run123456} shows the plots of individual energy 
fluxes from Run1-Run3; the bottom 
panel give their counterparts for Run4-Run6. Here,
$\Pi_{u>}^{u<}$ is shown in red, $\Pi_{u>}
^{b<}$ in blue,   $\Pi_{b>}^{b<}$ in magenta,  $\Pi_{b>}
^{u<}$ in orange, $\Pi_{b>}^{b<}$(Hall) in green, and  
$\Pi_{b>}^{j<}$(Hall) in purple. In [Run1-Run3], we 
observe that in $\Pi_{u}$, $\Pi_{u>}^{b<}$ dominates over $\Pi_{u>}
^{u<}$ for Run1 (MHD case); in 3D HMHD 
turbulence $\Pi_{u>}^{u<}$ dominates over $
\Pi_{u>}^{b<}$ and continues to dominate with the 
increase of $d_{i}$, particularly, in the second inertial 
region; furthermore, the behavior of $\Pi_{u>}^{b<}$ is unchanged, relative to 3D MHD (Run1), for 3D HMHD 
(Run2 and Run3). In $\Pi_{b}$, we observe that $\Pi_{b>}^{u<}$  
dominates over $\Pi_{b>}^{b<}$ for both 
3D MHD (Run1) and 3D HMHD (Run2 and Run3), but 
their amplitudes decrease at all wavenumbers as we 
increase $d_{i}$. $\Pi_{Hall}$: both $\Pi_{b>}
^{b<}$(Hall) and $\Pi_{u>}^{j<}$(Hall) are comparable, 
particularly in the second inertial region; however,  
$\Pi_{b>}^{j<}$(Hall) is comparatively 
more extended at higher wavenumbers; and their amplitudes increase with $d_{i}$. 
\par
Now we turn to [Run4-Run6]: In $\Pi_{u}$, $\Pi_{u>}^{u<}$ dominates over
$\Pi_{u>}^{b<}$ for low $Pr_m$ 3D HMHD turbulence (Run4), whereas, for Run5 and Run6, $\Pi_{u>}^{b<}$ dominates over $\Pi_{u>}^{u<}$; the amplitudes of $\Pi_{u>}^{u<}$ and $\Pi_{u>}^{b<}$ depending upon $Pr_m$ and show the following trends:

\begin{figure*}[!htbp]
	\centering
    \includegraphics[width=1.6\columnwidth]{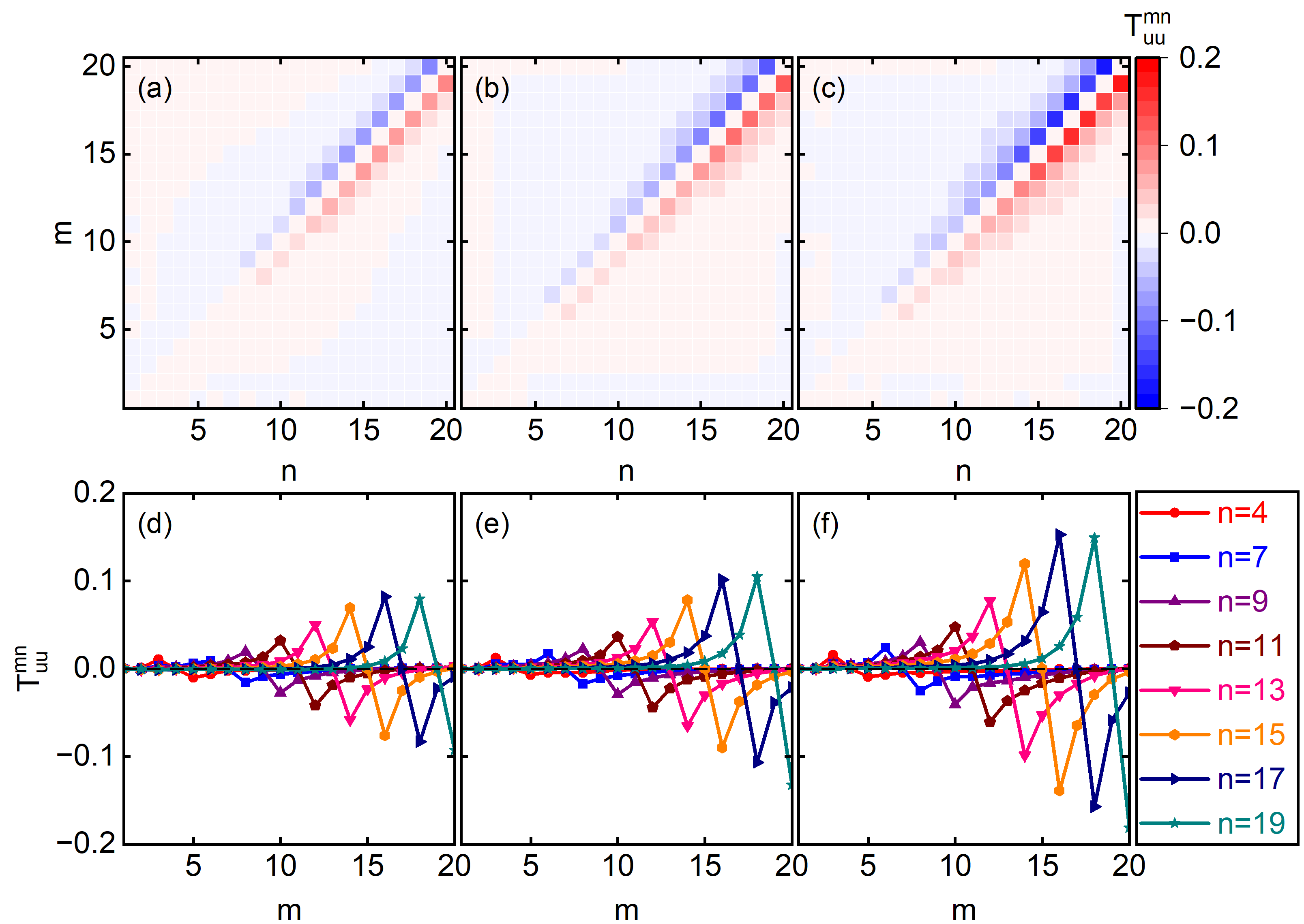}    
	\caption{\small Kinetic-to-kinetic $T^{mn}_{uu}$ energy transfer rate due to the non-linear term  $(\bm{u}\cdot\nabla)\bm{u}$ for [(a) and (d)] Run1, [(b) and (e)] Run2 and [(c) and (f)] Run3.}
    \label{fig:Tuu_Run123}
\end{figure*}

\begin{figure*}[!htbp]
	\centering
    \includegraphics[width=1.6\columnwidth]{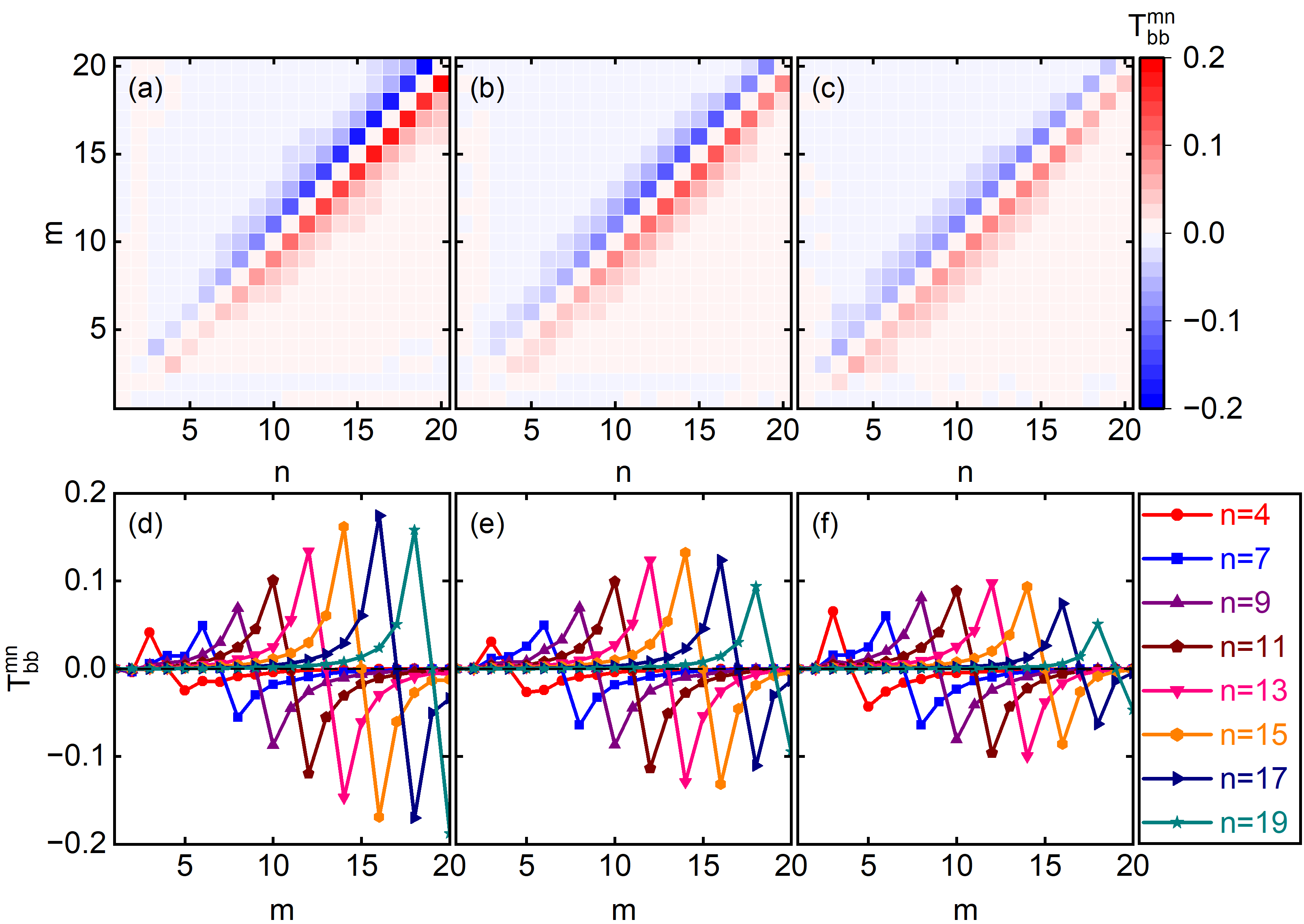}    
	\caption{\small Magnetic-to-magnetic $T^{mn}_{bb}$ energy transfer rate due to the non-linear term  $(\bm{u}\cdot\nabla)\bm{b}$ for [(a) and (d)] Run1, [(b) and (e)] Run2 and [(c) and (f)] Run3.}
        \label{fig:Tbb_Run123}
\end{figure*}
\begin{itemize}
\item $(\Pi_{u>}^{u<})^{Run4} > (\Pi_{u>}^{u<})^{Run5} > (\Pi_{u>}^{u<})^{Run6}$; 
\item and$(\Pi_{u>}^{b<})^{Run4} < (\Pi_{u>}^{b<})^{Run5} < (\Pi_{u>}^{b<})^{Run6}$.
\end{itemize}
In $\Pi_{b}$,  $ \Pi_{b>}^{b<}$ dominates over $\Pi_{b>}^{u<}$ in Run4; by contrast, in Run5 and Run6 $\Pi_{b>}^{u<}$ dominates over $\Pi_{b>}^{b<}$, and the trends of their amplitudes are as follows:
\begin{itemize}
\item $(\Pi_{b>}^{b<})^{Run4} > (\Pi_{b>}^{b<})^{Run5} > (\Pi_{b>}^{b<})^{Run6}$;
\item and the amplitude for $(\Pi_{b>}^{u<}$ is increased in Run5 in comparison to Run4 and Run6.
\end{itemize}
The amplitudes for both $(\Pi_{b>}^{b<})$(Hall) and $(\Pi_{b>}^{j<})$(Hall) in $\Pi_{Hall}$ are suppressed significantly for Run4 and Run6 in comparison to Run5; however, both of them are almost negiligible in Run4.

\subsection{Shell-to-shell Transfers}
\label{subsec:shell-shell}

Now, we present our results on the shell-to-shell energy transfer calculation for Run1-Run6. The Fourier space is partitioned into $N_s=20$ spherical shells $s_n=[k_n,k_{n+1}]$, where $k_n=2^{(n+8)/4}$. The first three shells are defined separately to ensure sufficient mode coverage: $s_1=[0,2]$, $s_2=[2,4]$, and $s_3=[4,8]$. The remaining shells follow the geometric law, so that the $m^{th}$ shell spans $[k_m, k_{m+1}]$. 

\begin{figure*}[!htbp]
	\centering
    \includegraphics[width=1.6\columnwidth]{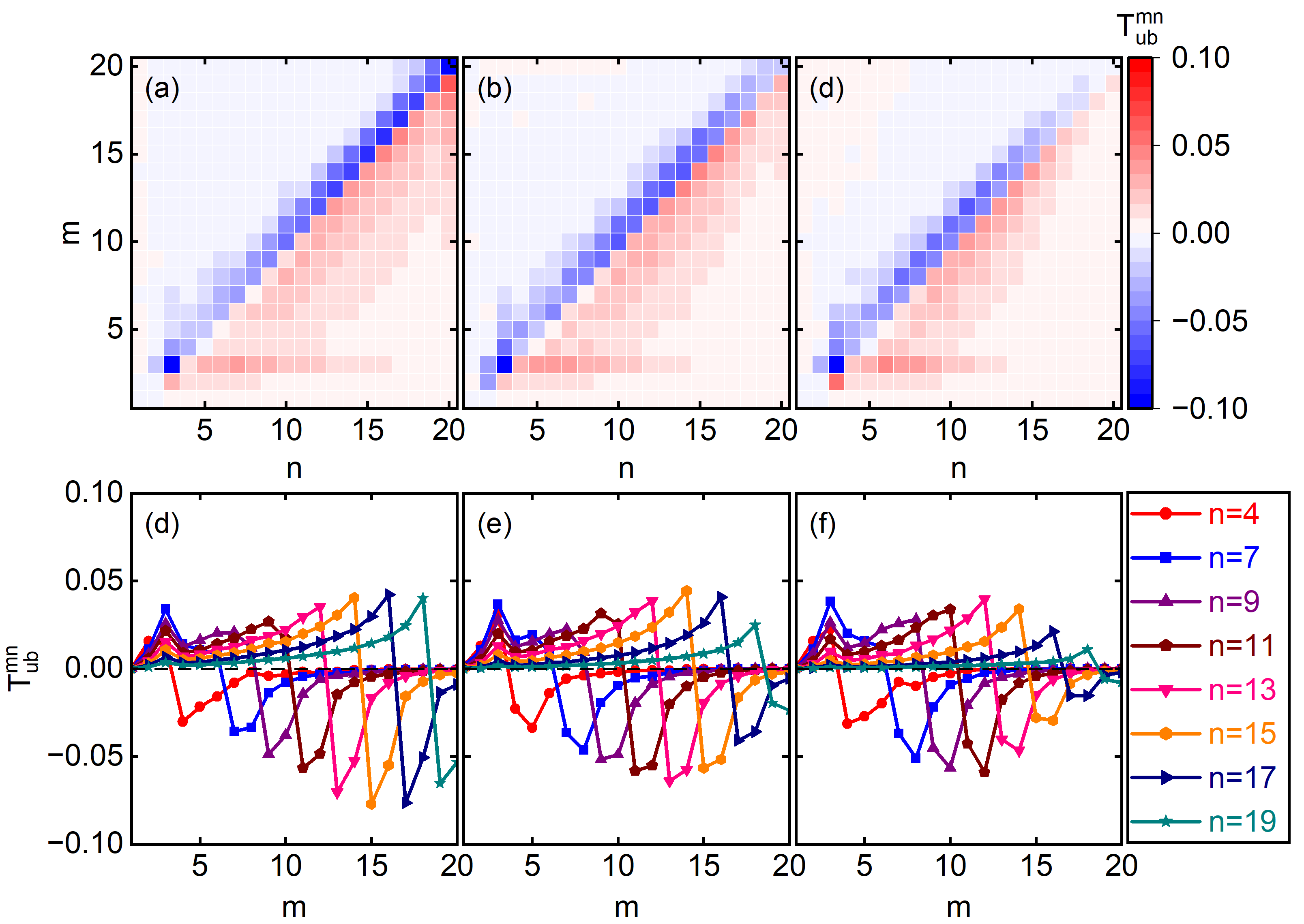}    
	\caption{\small Kinetic-to-magnetic $T^{mn}_{ub}$ energy transfer rate because of  the non-linear term  $(\bm{b}\cdot\nabla)\bm{u}$ for [(a) and (d)] Run1, [(b) and (e)] Run2 and [(c) and (f)] Run3.}
    \label{fig:Tub_Run123}
\end{figure*}

Transfers of kinetic-to-kinetic, kinetic-to-magnetic, magnetic-to-magnetic, and magnetic-to-kinetic are local when the shells involved are close in wavenumber space and non-local when they are widely separated. Figures \ref{fig:Tuu_Run123} (a), (b), and (c) illustrate the kinetic-to-kinetic energy transfers for Run1, Run2, and Run3, respectively. In the color scheme adopted here, the transition from white to red for a given square indicates that the giver shell m transfers energy to the receiver shell n ($T^{mn}>0$) while the transition from white to blue indicates that the giver shell $m$ receives energy from shell $n$ ($T_{mn}^<0$). For a given shell $m$, we observe that $T_{uu}^{mn}<0$ for $m>n$ and $T_{uu}^{mn}>0$ for $m<n$ [Fig. \ref{fig:Tuu_Run123} (a)], which is consistent with a forward energy cascade from large to small length scales. Furthermore, $T^{mn}_{uu}=0$ for $m=n$, confirming the absence of self-interactions - i.e., interactions between receiver and giver modes within the same shell do not alter the shell's net energy. The color distribution reveals that only shells in the vicinity of the diagonal ($m\simeq n$) interact appreciably, with interactions rapidly decaying away from it, indicating negligible non-local transfers. Note that large and small length scales respectively correspond to small and large wavenumbers. \\
To quantify the relative contributions of local and non-local energy transfer, we present line plots of the transfer rate as a function of the giver shell $m$ for several receiver shells $n$. Figures~\ref{fig:Tuu_Run123} (d), (e), and (f) demonstrate that, for a given receiver shell $n$, the transfer amplitude peaks at nearby shells and diminishes rapidly with increasing shell separation, confirming the predominantly local nature of $T^{mn}_{uu}$. We observe that the transfer amplitudes increase with the receiver shell index. Furthermore, for Run3, energy transfer is enhanced at high wavenumbers relative to both Run1 ($d_i = 0.0$) and Run2 ($d_i = 0.025$). Therefore, the results of Fig.~\ref{fig:Tuu_Run123} confirm that kinetic-to-kinetic energy transfers are strictly forward, proceeding from low to high wavenumbers in both 3D MHD and 3D HMHD systems. 
Figure \ref{fig:Tbb_Run123} illustrates the magnetic-to-magnetic energy transfers for Run1, Run2, and Run3. Figures \ref{fig:Tbb_Run123} (a), (b), and (c) show the forward energy cascade from large to small length scales of $T_{bb}^{mn}$ for Run1, Run2, and Run3, which is consistent with our results for $T_{mn}^{uu}$; also, there is no self-interaction  for $T^{mn}_{bb}$.\\ 
Figures \ref{fig:Tbb_Run123} (d), (e), and (f) demonstrate the local nature of energy transfer for $T^{mn}_{bb}$; this is similar to $T^{mn}_{uu}$. Note that the transfer amplitudes increase with the receiver shell index $n$. Furthermore, for Run1, the energy transfer rate is higher, at large wavenumbers, than in both Run2 ($d_i = 0.025$) and Run3 ($d_i = 0.05$). Thus, Fig.~\ref{fig:Tbb_Run123}) demonstrates that magnetic-to-magnetic energy transfers are strictly forward and local, as they are for $T_{uu}^{mn}$. We find enhanced transfer rates, $T_{uu}^{mn}$ and $T_{bb}^{mn}$, for Run3 (3D HMHD) and Run1 (3D 
MHD), respectively.\\
We next examine the cross-field energy transfers between velocity and magnetic fields. Figures \ref{fig:Tub_Run123} (a), (b), and (c) reveal that $T_{ub}^{mn}<0$ for $m=n$, implying that within the same shell the velocity field gains energy from the magnetic field. However, $T^{mn}_{ub} < 0$ for $m>n$ and $T^{mn}_{ub} > 0$ for $m<n$, indicating that the velocity field receives energy from lower-wavenumber magnetic modes and subsequently redistributes it to higher-wavenumber magnetic modes. The color distribution in this plot suggests that the cross-field transfer is forward and more non-local in character than the same-field transfer. Figures \ref{fig:Tub_Run123} (d), (e), and (f) further quantify these contributions: for a given receiver shell, the velocity field gains energy from the magnetic field primarily through local transfer (nearby or same-shell), whereas it loses energy to the magnetic field through both local and non-local transfer spanning nearby and distant shells.

  \begin{figure*}[!htbp]
	\centering
    \includegraphics[width=1.6\columnwidth]{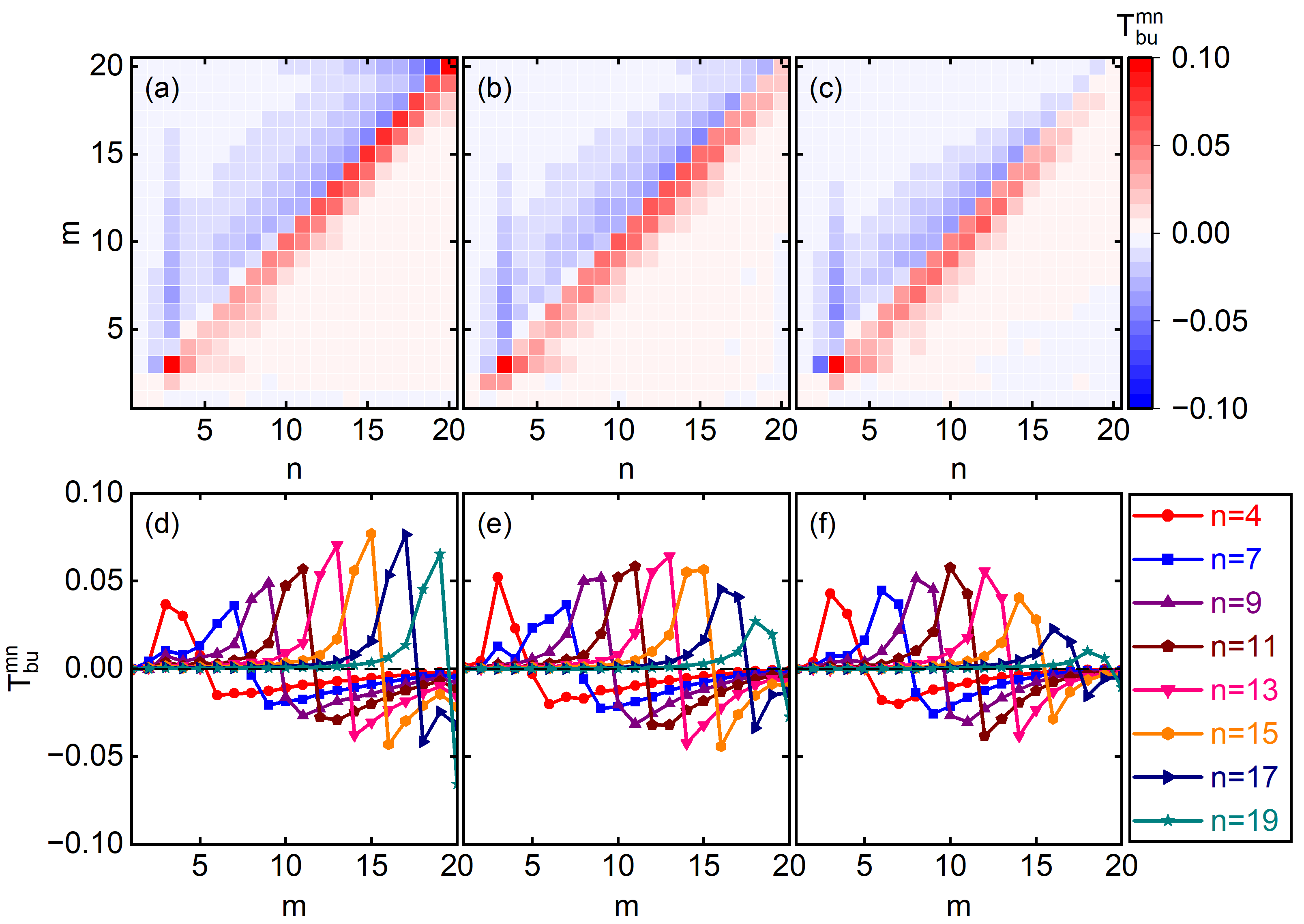}
	\caption{\small Magnetic-to-kinetic $T^{mn}_{bu}$ energy transfer rate because of the non-linear term  $(\bm{b}\cdot\nabla)\bm{b}$ for [(a) and (d)] Run1, [(b) and (e)] Run2 and [(c) and (f)] Run3.}
    \label{fig:Tbu_Run123}    
\end{figure*}
 Similarly, Figs. \ref{fig:Tbu_Run123} (a), (b), and (c) show the magnetic-to-kinetic $T^{mn}_{bu}$ energy transfer, which also indicates the local and non-local forward energy transfer of magnetic-to-kinetic from small wavenumbers to large wavenumbers. Figures \ref{fig:Tbu_Run123} (d), (e), and (f) show the strength of local and non-local energy transfers. The velocity field for a given shell receives energy from the magnetic field from nearby or the same shell (i.e., only local transfers are significant) and loses energy to the magnetic field in nearby and distant shells (local and non-local transfers). These results demonstrate that the cross-field energy transfer mechanisms in Hall-MHD turbulence remain qualitatively similar to those in standard MHD turbulence. These transfer rates are also not affected by the strength of the Hall term.
 \par
\begin{figure*}[!htbp]
	\centering
    \includegraphics[width=1.6\columnwidth]{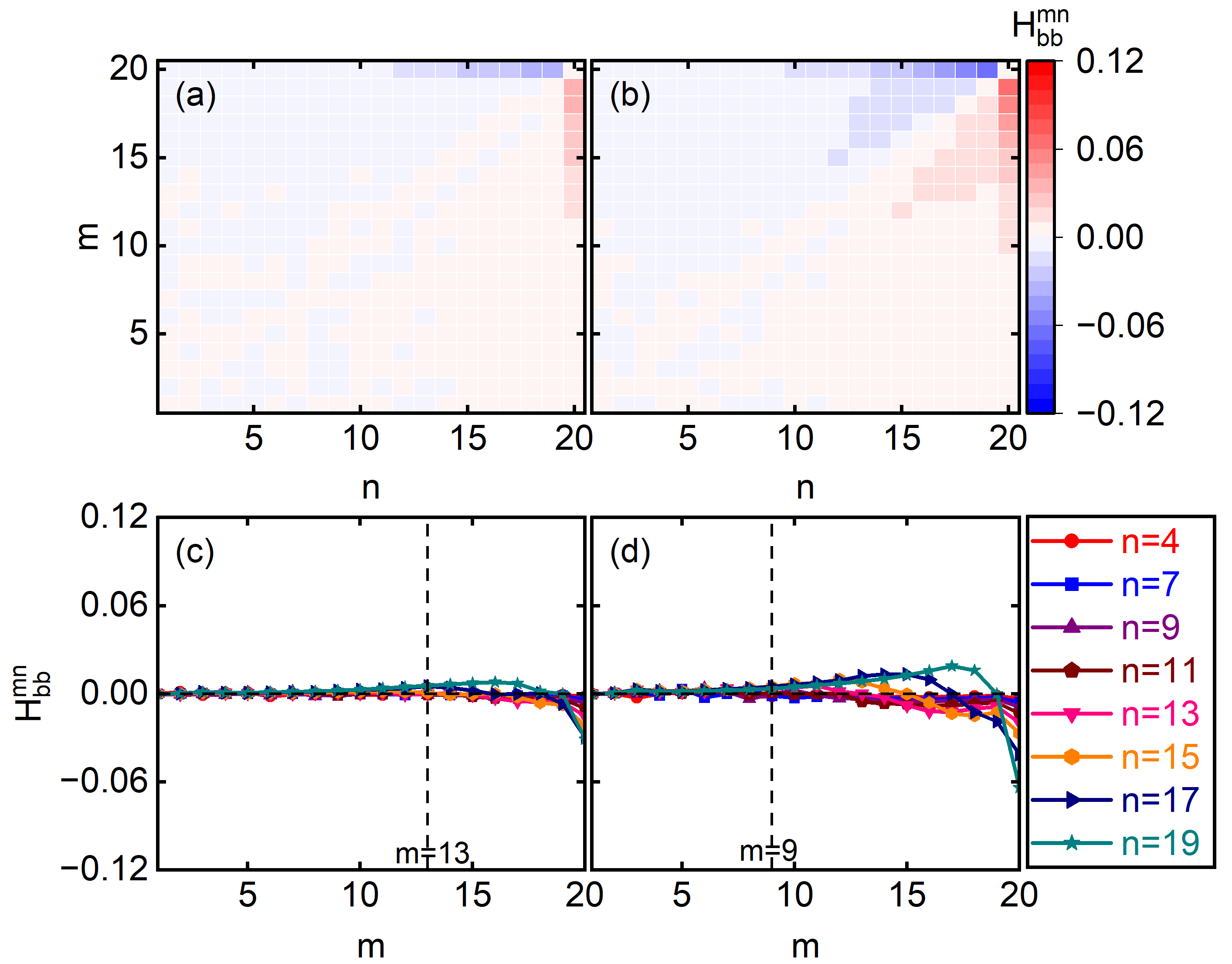}    
	\caption{\small Magnetic-to-magnetic $H^{mn}_{bb}$ energy transfer rate because of the Hall term  $(\bm{j}\cdot\nabla)\bm{b}$ for [(a) and (c)] Run2 and [(b) and (d)] Run3.}
    \label{fig:Hbb_Run23}
\end{figure*}
\begin{figure*}[!htbp]
	\centering
    \includegraphics[width=1.6\columnwidth]{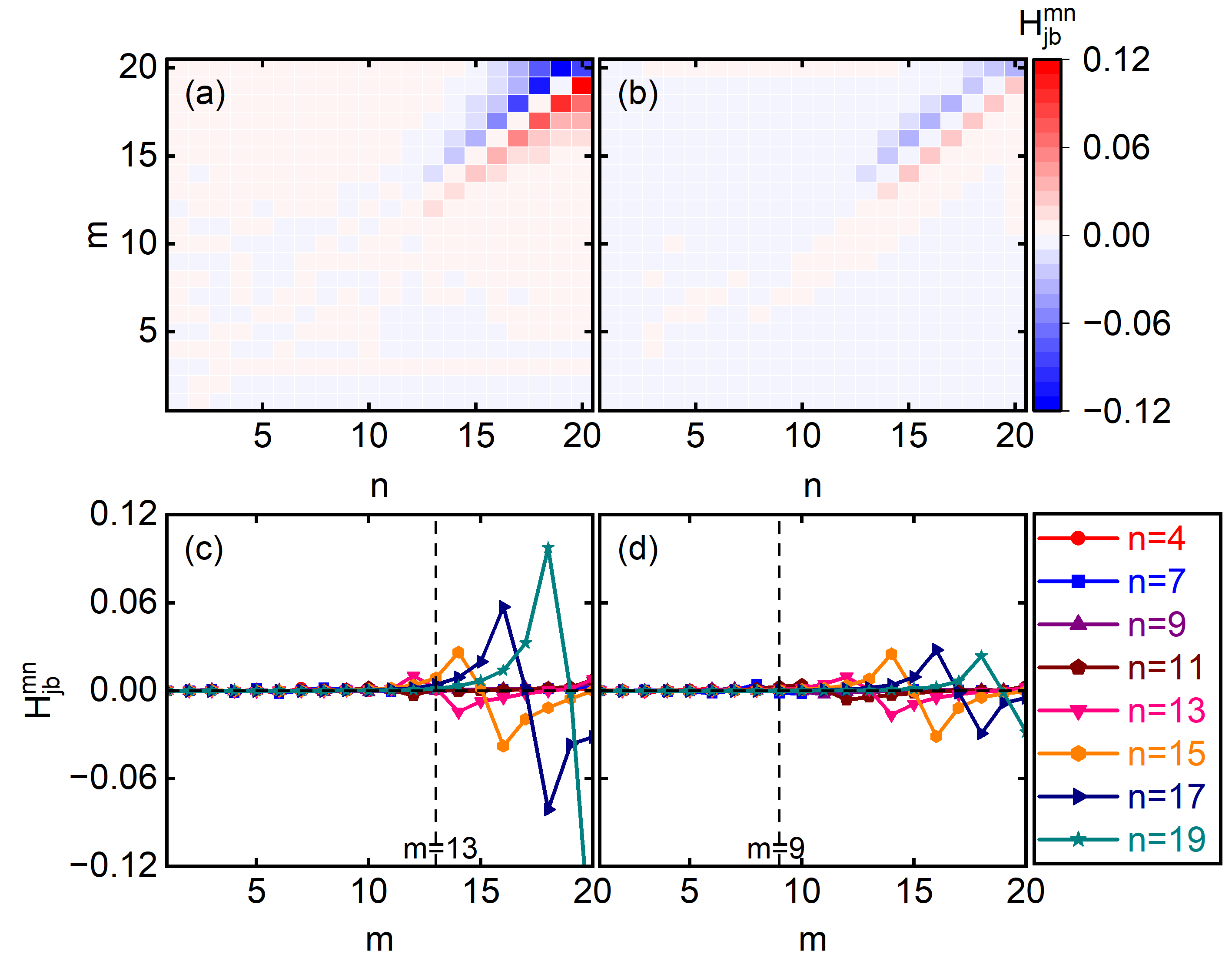}    
	\caption{\small Current density-to-magnetic $H^{mn}_{jb}$ energy transfer rate because of the Hall term  $(\bm{b}\cdot\nabla)\bm{j}$ for [(a) and (c)] Run2 and [(b) and (d)] Run3.} 
    \label{fig:Hjb_Run23}
\end{figure*}

 As discussed previously, the Hall term introduces an additional nonlinear contribution to energy transfer in Hall-MHD turbulence. Let $H^{mn}_{bb}$ denote the magnetic-to-magnetic energy transfer mediated by the current density field; and define $H^{mn}_{jb}$ to be the current-density-to-magnetic energy transfer mediated by the magnetic field. Figure~\ref{fig:Hbb_Run23} presents the energy transfers governed by the Hall term [see Eqs.~\eqref{Eq:23} and \eqref{Eq:24}]. The magnetic-to-magnetic transfer $H^{mn}_{bb}$ remains negligible and exhibits fluctuations up until the receiver shells $n = 15$ and $n=11$ for Run2 and Run3, respectively [see Figs.~\ref{fig:Hbb_Run23} (a) and (b)]; beyond this, it develops signatures of both local and non-local forward transfers, becoming increasingly prominent at small length scales. This is corroborated by the line plots in Figs. \ref{fig:Hbb_Run23} (c) and (d), which confirm a forward magnetic-to-magnetic transfer at small length scales, albeit with a weak amplitude. Here, the vertical dashed line indicates the shell number that contains the wavenumber, corresponding to the ion-inertial length $k_i\sim 1/d_i$ [$k_i=40$(Run2) and $20$ (Run3)]. These observations suggest that $H^{mn}_{bb}$ is essentially confined to the small length scales dynamics and does not play a role in large length scale energy redistribution. The reason we do not observe any inverse transfer is that flux rates in the second-inertial region are also influenced by the non-linear terms [see figs.\ref{fig:flux_rates_Run123456} and \ref{fig:termwise_flux_rates_Run123456}]. Figures \ref{fig:Hjb_Run23} (a) and (b) illustrate the current-density-to-magnetic energy transfer $H^{mn}_{jb}$, which remains relatively weak up to shell number $n = 11$, after which its amplitude grows progressively with increasing wavenumber. At small length scales (or higher wavenumber), $H^{mn}_{jb}$ exhibits strong local and forward transfer characteristics. These results indicate that the energy transfer driven by the Hall term is predominantly confined to small length scales dynamics. Its influence on large-length-scale energy transfer is negligible, and its most significant contributions arise in the high-wavenumber regime, where the Hall term becomes relevant.

\par
\begin{figure*}[!htbp]
	\centering
    \includegraphics[width=1.6\columnwidth]{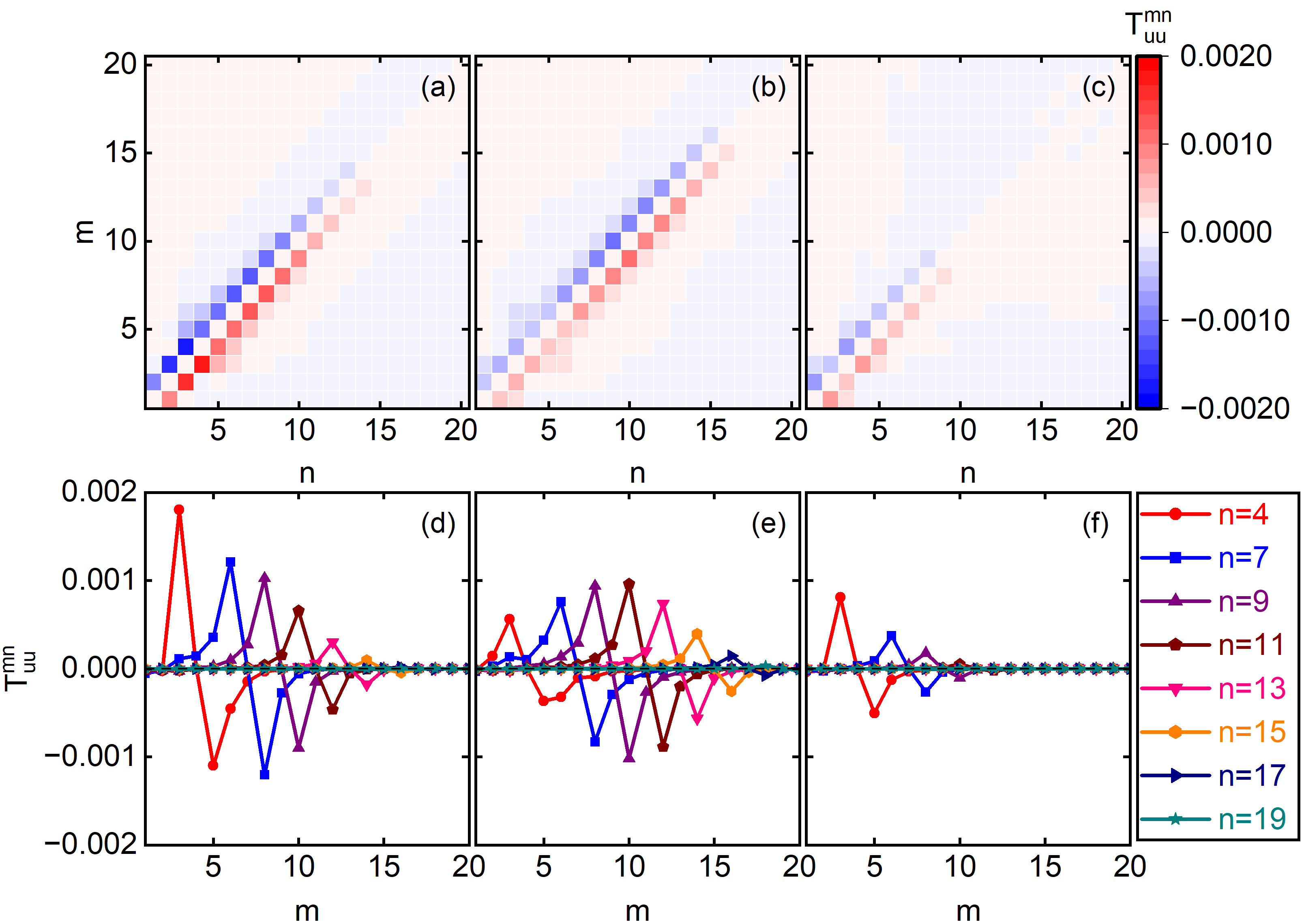}    
	\caption{\small kinetic-to-kinetic $T^{mn}_{uu}$ energy transfer rate because of the non-linear term  $(\bm{u}\cdot\nabla)\bm{u}$ for [(a) and (d)] Run4, [(b) and (e)] Run5 and [(c) and (f)] Run6.}
    \label{fig:Tuu_Run456}
\end{figure*}
\begin{figure*}[!htbp]
	\centering
    \includegraphics[width=1.6\columnwidth]{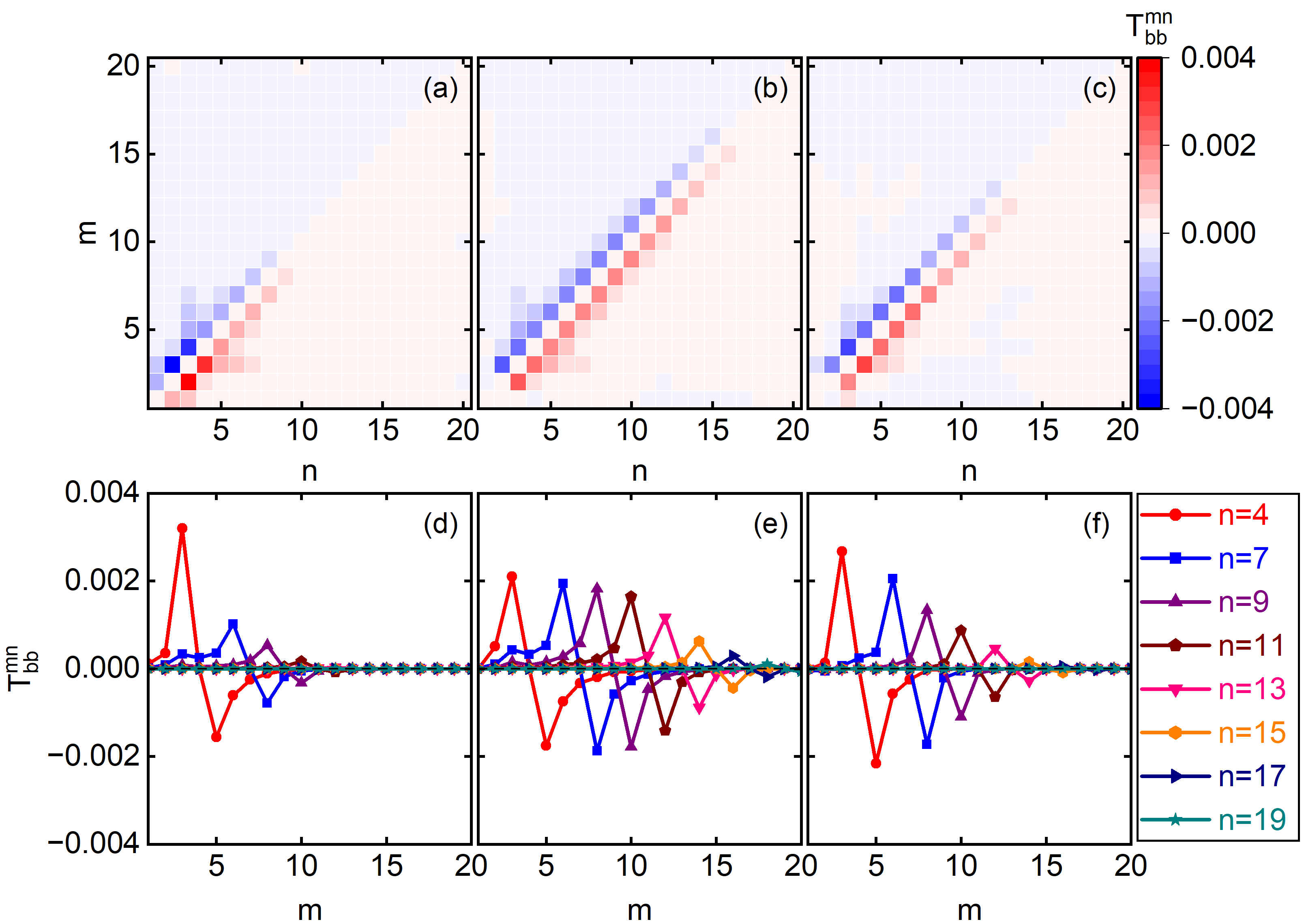}    
	\caption{\small Magnetic-to-magnetic $T^{mn}_{bb}$ energy transfer rate because of the non-linear term  $(\bm{u}\cdot\nabla)\bm{b}$ for [(a) and (d)] Run4, [(b) and (e)] Run5 and [(c) and (f)] Run6.} 
    \label{fig:Tbb_Run456} 
\end{figure*}

We provide a systematic comparison of the shell-to-shell energy transfer results for Run4, Run5, and Run6, corresponding to $Pr_m=0.1$, $1$, and $10$, respectively. Not only do these runs differ in their magnetic Prandtl numbers, but they also employ initial energy spectra [see Fig.~\ref{fig:initialenrgyspectra}] that are distinct from those used in [Run1-Run3]. Figure \ref{fig:Tuu_Run456} presents the shell-to-shell kinetic-to-kinetic energy transfer $T^{mn}_{uu}$ for Run4, Run5, and Run6. At $Pr_m = 1.0$ [Run4], $T^{mn}_{uu}$ is predominantly local and of the forward-transfer type; 
and it remains visible across the full range of wavenumber scales [Figs. \ref{fig:Tuu_Run456} (b) and (e)]. For $Pr_m = 0.1$ [Figs. \ref{fig:Tuu_Run456} (a) and (d)] and $Pr_m = 10$ [Figs. \ref{fig:Tuu_Run456} (c) and (f)], this transfer retains its local and forward-transfer character but becomes increasingly difficult to observe at small length scales, because of the enhanced magnetic diffusivity and kinematic viscosity at low and high Prandtl numbers, respectively. At high magnetic Prandtl numbers, the energy transfer driven by the convective term $(\bm{u} \cdot \nabla)\bm{u})$ remains local and of the forward-transfer
type, but its amplitude is noticeably reduced. We find, in consonance with the earlier 3D HMHD results, that $T^{mn}_{uu}=0$ for $m=n$ and all the values of $Pr_m$ we consider; i.e., intra-shell interactions do not contribute to net energy exchange between shells.  Figure~\ref{fig:Tuu_Run456} shows that, for a fixed receiver shell, the nearest-neighbor shells $n = m \pm 1$ serve as the dominant contributors to energy transfer, which is a hallmark 
of local cascades.\\
In Fig. \ref{fig:Tbb_Run456} we examine, for Run4, Run5, and Run6, the magnetic-to-magnetic energy transfer $T^{mn}_{bb}$, driven by the nonlinear term $(\bm{u} \cdot \nabla)\bm{b})$ in the induction equation ~\eqref{eq:mag}. For all values of $Pr_m$, we show that $T^{mn}_{bb}$ is predominantly local and exhibits a forward energy cascade. $T^{mn}_{bb}$ is confined to a narrow range of wavenumbers at low $Pr_m$ because of the dominance of the
magnetic diffusivity over kinematic viscosity. The corresponding line plots in Figs. \ref{fig:Tbb_Run456} (d), (e), and (f) are consistent with the density plots, exhibiting the same local and forward-transfer signatures. Furthermore, for all the values of $Pr_m$ we study, the peak-transfer amplitude for a given receiver shell decreases with increasing shell number, as we have observed for $T^{mn}_{uu}$.

\par
\begin{figure*}[!htbp]
	\centering
    \includegraphics[width=1.6\columnwidth]{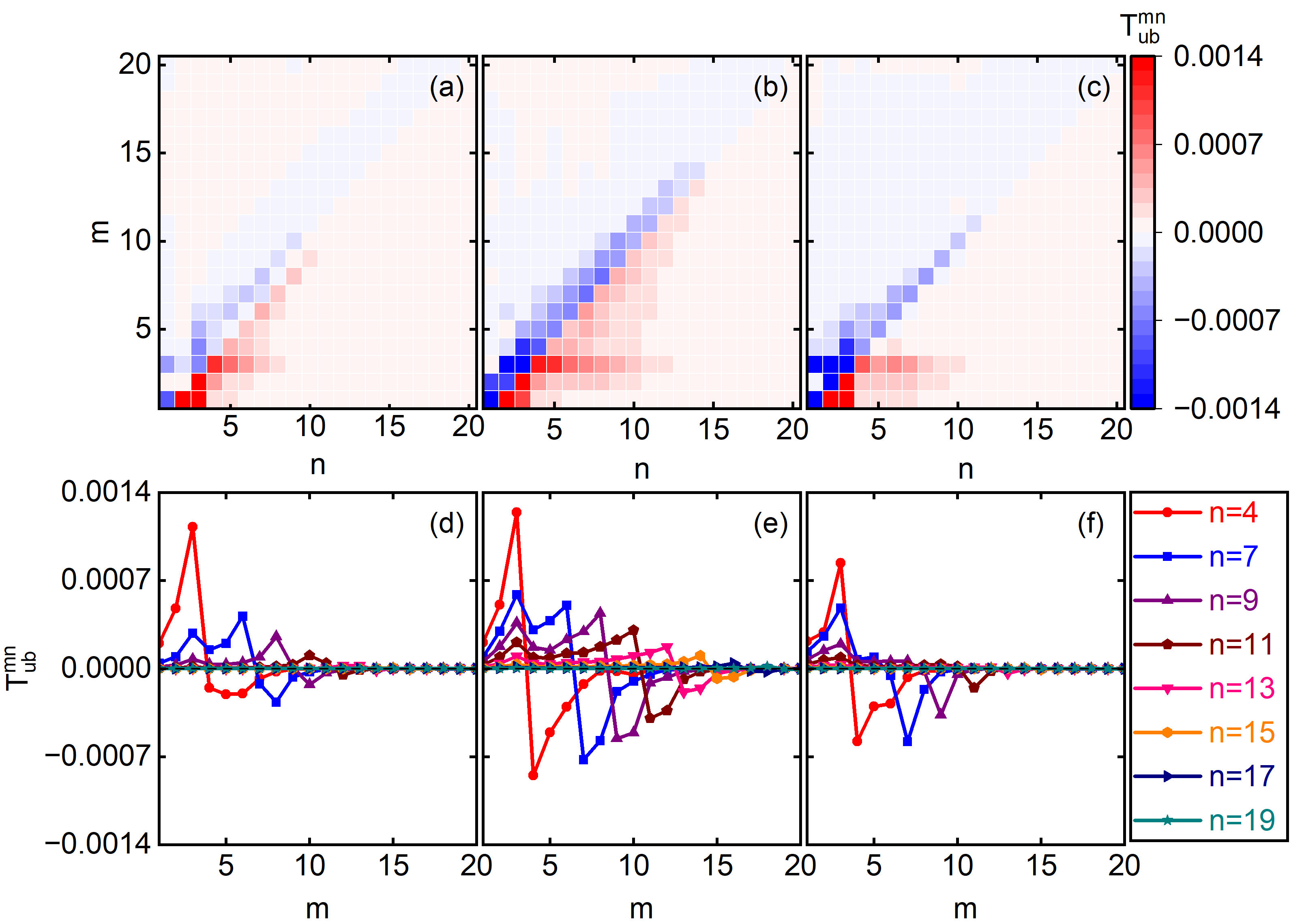}    
	\caption{\small kinetic-to-magnetic $T^{mn}_{ub}$ energy transfer rate because of the non-linear term  $(\bm{b}\cdot\nabla)\bm{u}$ for [(a) and (d)] Run4, [(b) and (e)] Run5 and [(c) and (f)] Run6.}
    \label{fig:Tub:Run456}
\end{figure*}

We now present the  kinetic-to-magnetic energy transfer $T^{mn}_{ub}$, associated with the nonlinear term $(\bm{b} \cdot \nabla)\bm{u})$ in the momentum equation~\eqref{eq:vel}, for Run4, Run5, and Run6, as shown in Fig. \ref{fig:Tub:Run456}. At $Pr_m=1.0$ [Figs. \ref{fig:Tub:Run456} (a) and (d)], $T^{mn}_{ub}$ is negative for $m\geq n$ and positive for $m \ll n$, consistent with the behavior observed in Run1-Run3 (Fig. \ref{fig:Tub_Run123}). This pattern indicates that the velocity field receives energy from the magnetic field within the same and lower-wavenumber shells, and subsequently transfers it to the magnetic field at higher-wavenumber shells. However, the transfer rates at higher shell numbers are comparatively weaker than those observed in Run1, Run2, and Run3, likely due to the different initial energy spectra used in these runs. The density plot of $T^{mn}_{ub}$ [Fig. \ref{fig:Tub:Run456} (b) and (c)] exhibit the clear signatures of non-local forward energy transfer from the large length scale velocity field to the small length scale magnetic field. At $Pr_m = 0.1$ and $Pr_m = 10$, the overall transfer characteristics remain qualitatively similar. However, the transfer amplitudes are noticeably suppressed, which is a consequence of enhanced magnetic diffusivity and kinematic viscosity at low and high Prandtl numbers, respectively. This suppression is consistent with the trends observed for $T^{mn}_{uu}$ and $T^{mn}_{bb}$ at comparable Prandtl numbers. The line plots in Figs. \ref{fig:Tub:Run456} (d), (e), and (f) confirm that the transfer amplitude decreases monotonically with increasing receiver shell number for all three runs.
\par
\begin{figure*}[!htbp]
	\centering
    \includegraphics[width=1.6\columnwidth]{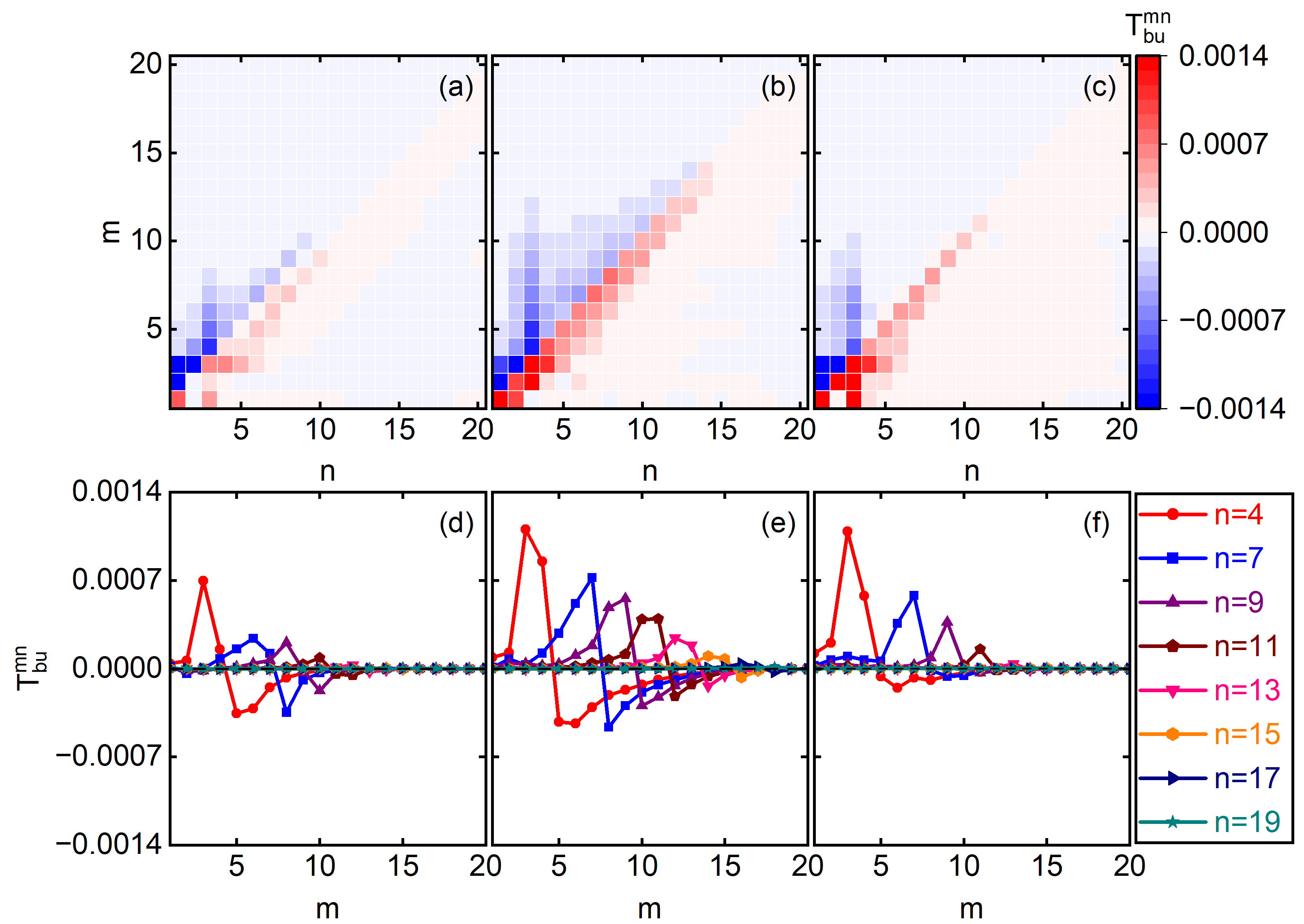}    
	\caption{\small Magnetic-to-kinetic $T^{mn}_{bu}$ energy transfer rate because of the non-linear term  $(\bm{b}\cdot\nabla)\bm{b}$ for [(a) and (d)] Run4, [(b) and (e)] Run5 and [(c) and (f)] Run6.}
    \label{fig:Tbu:Run456}    
\end{figure*}

Figure \ref{fig:Tbu:Run456} presents the magnetic-to-kinetic energy transfer $T^{mn}_{bu}$, associated with the nonlinear term $(\bm{b} \cdot \nabla)\bm{b}$ in the momentum equation~\eqref{eq:vel}, for Run4, Run5, and Run6. Similar to $T^{mn}_{ub}$, the transfer $T^{mn}_{bu}$ exhibits both local and non-local forward characteristics. Specifically, $T^{mn}_{bu}$ is negative for $m \gg n$ and positive for $m<n$, indicating that the magnetic field gains energy from the velocity field for low-wavenumber shells, and $T^{mn}_{bu}>0$ for $n=m$.

\par
 The negative component of $T^{mn}_{bu}$ is more non-local than its positive counterpart, suggesting that overall transfers are both local and non-local. At $Pr_m = 0.1$ and $Pr_m = 10$, the overall transfer characteristics remain qualitatively consistent with those at $Pr_m = 1.0$, although the active transfer range is narrowed to a limited wavenumber range because of the strong magnetic diffusivity and kinematic viscosity at low and high Prandtl numbers, respectively. This behavior is analogous to that observed for $T^{mn}_{ub}$ under similar conditions. Furthermore, we quantify the local and non-local contributions [Figs. \ref{fig:Tbu:Run456} (d), (e), and (f)] by presenting the transfer rates $T^{mn}_{ub}$ as functions of the giver shell for several fixed values of the receiver shells for all three values of $Pr_m$. These line plots confirm that, for $Pr_m = 0.1$ and $Pr_m = 10$, the transfer amplitude diminishes with increasing shell number, and the signatures of non-local transfers are progressively suppressed.
\par

\begin{figure*}[!htbp]
	\centering
    \includegraphics[width=1.6\columnwidth]{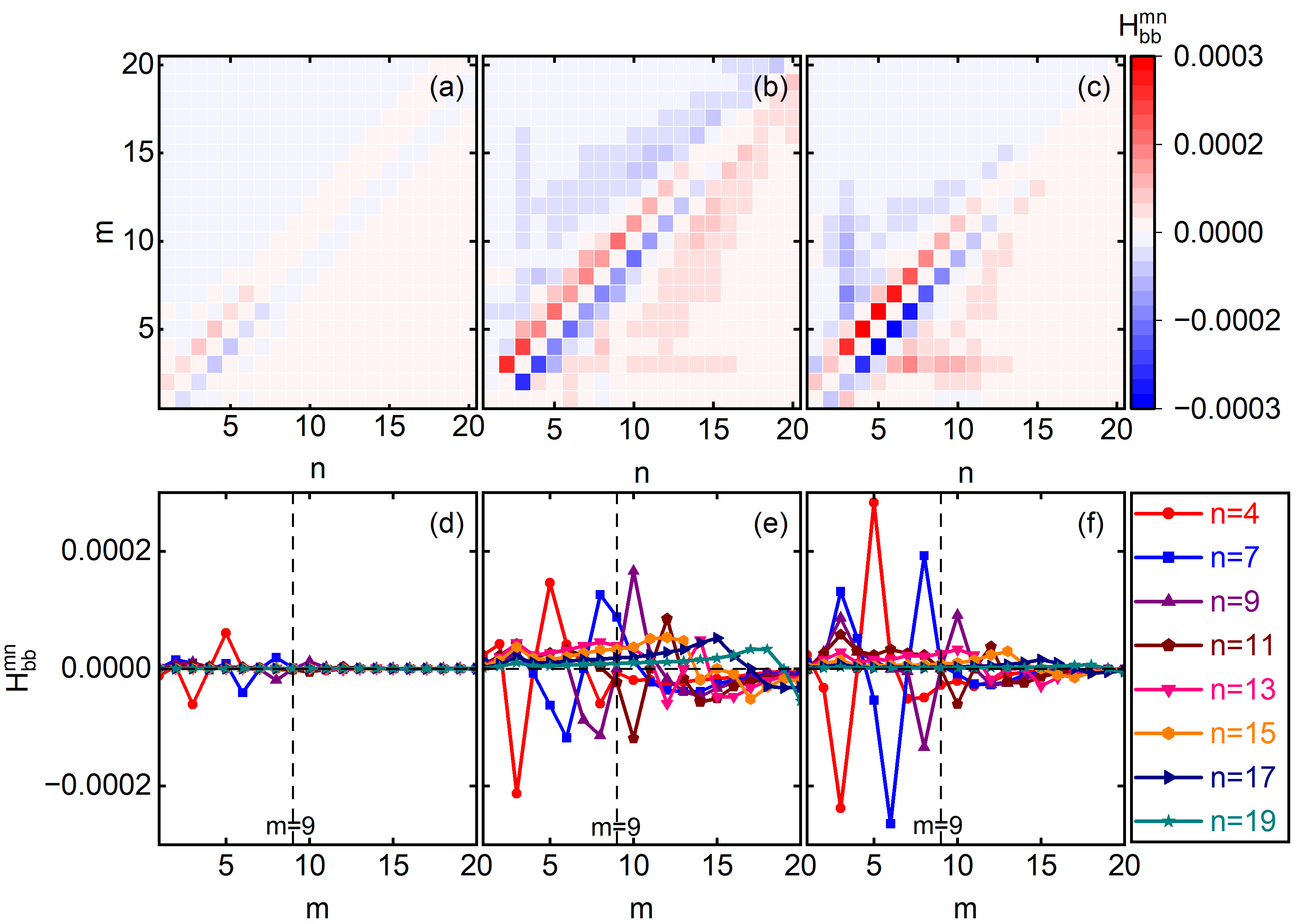}    
	\caption{\small Magnetic-to-magnetic $H^{mn}_{bb}$ energy transfer rate because of the Hall term  $(\bm{j}\cdot\nabla)\bm{b}$ for [(a) and (d)] Run4, [(b) and (e)] Run5 and [(c) and (f)] Run6.} 
    \label{fig:Hbb_Run456}
\end{figure*}

Consider now the energy-transfer contributions from the Hall term $d_i \nabla \times (\bm{j} \times \bm{b})$ for Run4, Run5, and Run6. As outlined in Sec.~\ref{subsec:fluxrates}, this term can be decomposed into two distinct nonlinear contributions, whose corresponding shell-to-shell energy transfer formulations are given by Eqs.~\eqref{Eq:23} and \eqref{Eq:24}.
Figure \ref{fig:Hbb_Run456} presents the magnetic-to-magnetic energy transfer $H^{mn}_{bb}$, driven by the nonlinear term $(\bm{j} \cdot \nabla)\bm{b})$, for Run4, Run5, and Run6. $H^{mn}_{bb}$ vanishes for $m = n$, confirming that there is no net magnetic-to-magnetic energy exchange between modes residing within the same shell from the Hall term. In the inertial range ($k\ll k_{i}$), $H^{mn}_{bb}$ is negative for receiver shells $n>m$ and positive for $n<m$ in the vicinity of the diagonal; this is indicative of a local inverse transfer of magnetic energy from small to large length scales.  Away from the diagonal, $H^{mn}_{bb}$ becomes positive for $n \gg m$ and negative for $m \gg n$; this points to a non-local forward transfer of magnetic energy from large to small length scales.

\begin{figure*}[!htbp]
	\centering
    \includegraphics[width=1.6\columnwidth]{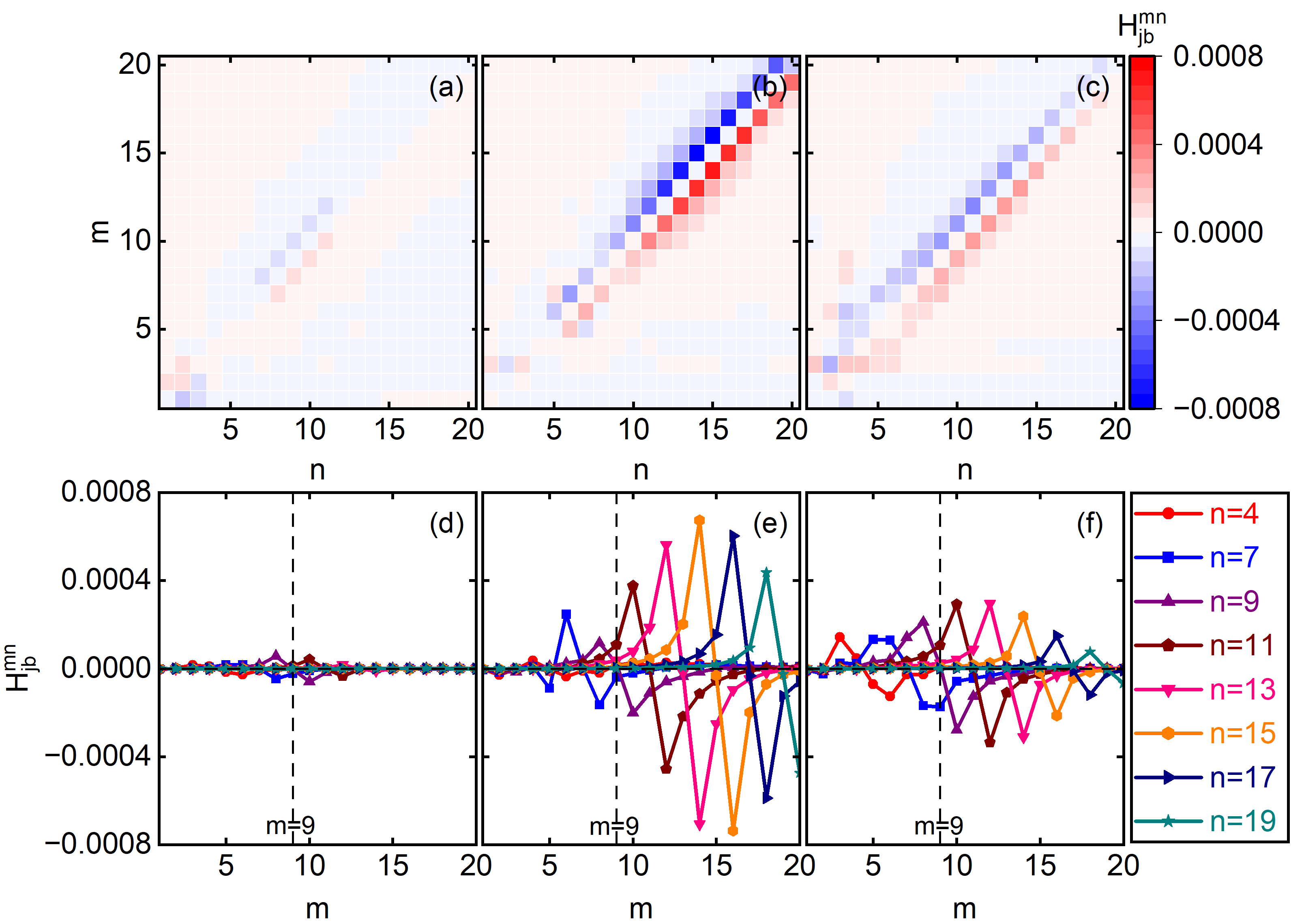}    
	\caption{\small Current density-to-magnetic $H^{mn}_{jb}$ energy transfer rate because of the Hall term  $(\bm{b}\cdot\nabla)\bm{j}$ for [(a) and (d)] Run4, [(b) and (e)] Run5 and [(c) and (f)] Run6.}
    \label{fig:Hjb_Run456}    
\end{figure*}

\noindent
 However, this forward component is weaker in comparison to the local inverse transfer. At large shell numbers, $H^{mn}_{bb}$ exhibits a weak, local forward transfer, confined to shells near the diagonal. At $Pr_m = 0.1$ [Fig. \ref{fig:Hbb_Run456} (a)], we observe local inverse transfers; however, the non-local forward transfer is suppressed relative to its counterpart for $Pr_m = 1.0$. At $Pr_m = 10$ [Fig. \ref{fig:Hbb_Run456}(c)], a local forward transfer and a weak non-local inverse transfer is evident; this is similar to that in Run5 ($Pr_m=1.0$). The line plots in Figs. \ref{fig:Hbb_Run456} (d), (e), and (f) present the transfer rates as a function of the giver shell, for fixed receiver shells, in Run4, Run5, and Run6, respectively. These plots indicate that $H^{mn}_{bb}$ exhibits the strongest transfer amplitude at $Pr_m = 10$, with the amplitude decreasing monotonically as the magnetic Prandtl number decreases. It is instructive to contrast these results with those for Run2 and Run3 [Fig.~\ref{fig:Hbb_Run23}], where the local inverse transfer in $H^{mn}_{bb}$ is absent, only local and non-local forward transfers are present, and these are confined exclusively to small length scales.
 \par
Figure \ref{fig:Hjb_Run456} presents, for Run4, Run5, and Run6, the current-density-to-magnetic energy transfer $H^{mn}_{jb}$, driven by the nonlinear term $(\bm{b} \cdot \nabla)\bm{j}$. At $Pr_m = 1.0$ [Fig. \ref{fig:Hjb_Run456} (b)], $H^{mn}_{jb}$ vanishes for $m=n$, confirming the absence of direct energy exchange between the current density and magnetic field within the same shell. The transfer exhibits oscillatory behavior up until shell number $n = 5$, beyond which a strong local forward transfer emerges and persists into the inertial and sub-inertial regions. At $Pr_m = 10$ [Fig. \ref{fig:Hjb_Run456} (c)], similar transfer characteristics are observed, spanning continuously from large to small length scales. At $Pr_m = 0.1$ [Fig. \ref{fig:Hjb_Run456} (a)], however, the transfer retains its local and forward character but is considerably weaker in amplitude compared to that higher values of $Pr_m$.
\par
The line plot in Fig.~\ref{fig:Hjb_Run456} (d) shows that, for a fixed receiver shell, the peak transfer amplitude increases with the giver shell number, reflecting the growing importance of small length scale interactions. Figures~\ref{fig:Hjb_Run456} (d) and (f) demonstrate, in addition, that the 
amplitude of $H^{mn}_{jb}$ is reduced in Run4 and Run6 relative to that in Run5; this is consistent with the suppression of the Hall-driven transfer at low and high Prandtl numbers. We conclude, therefore, that $H^{mn}_{jb}$ is most active in the Hall-term-transfer channel in the second-inertial region, insofar as it exhibits stronger and more persistent signatures than the other transfer components examined here.

\FloatBarrier

\section{Conclusions}
\label{sec:conclusion}

We have investigated the shell-to-shell energy transfer in 3D decaying Hall-MHD turbulence using shell decompositions of velocity- and magnetic-field data from our direct numerical simulations (DNS), which have been designed to study the dependence of these transfers on the ion-inertial length $d_{i}$ and the magnetic Prandtl number $Pr_{m}$. This has not been attempted hitherto. We have carried out a systematic comparison of energy transfer by different nonlinear terms for three values of $Pr_m$, namely, $Pr_m = 0.1$, 1.0, and 10 at the fixed value $d_{i}=0.05$; we have also explored two different values of $d_i$, namely, $d_{i}=0.025$ and 0.05, and, of course, $d_{i}=0.0$ (3D MHD) at the fixed value of $Pr_{m}=1.0$. Specifically, we have computed the shell-to-shell energy transfer functions $T^{mn}_{uu}$, $T^{mn}_{bb}$, $T^{mn}_{ub}$, $T^{mn}_{bu}$, $H^{mn}_{bb}$, and $H^{mn}_{jb}$ for all simulations that are summarized in Table \ref{table1}.
\par
We have demonstrated that the kinetic-to-kinetic ($T^{mn}_{uu}$) and magnetic-to-magnetic ($T^{mn}_{bb}$) energy transfers emerge as the dominant local forward-cascade channels in both 3D MHD and 3D HMHD turbulence for all the magnetic Prandtl numbers considered. Furthermore, we have obtained new insights into cross-field transfers, namely, kinetic-to-magnetic ($T^{mn}_{ub}$) and magnetic-to-kinetic ($T^{mn}_{bu}$) transfers: these exhibit clear signatures of non-local transfer at $Pr_m = 1.0$ in both MHD and Hall-MHD turbulence, in contrast to the like-field transfers. At $Pr_m = 0.1$ and $Pr_m = 10$, these non-local transfer signatures are substantially suppressed and remain observable only over a narrow wavenumber range, consistent with the increased influence of magnetic diffusivity and kinematic viscosity at these parameter values.
\par
We have obtained results for the magnetic-to-magnetic energy transfer driven by the Hall term, $H^{mn}_{bb}$, which exhibits a local inverse transfer of magnetic energy at large length scales, including the inertial region. In addition, it exhibits a weak non-local forward transfer across the large length scales and inertial region, and then crosses over to a local forward transfer at the small length scales. At $Pr_m = 0.1$, non-local forward transfers are suppressed relative to those at $Pr_m = 1.0$ and 10. At $Pr_m = 10$; and a strong local inverse transfer is observed; the non-local forward transfer remains comparable in magnitude to that for $Pr_m = 1.0$, but is absent at small length scales. The current-density-to-magnetic energy transfer $H^{mn}_{jb}$, also driven by the Hall term, remains weak at large length scales for $Pr_m = 1.0$ and $Pr_m = 10$; its amplitude grows progressively with shell number into the small length scale ranges. At $Pr_m = 0.1$, this transfer is significantly suppressed across all length scales.
\par
A recent study~\cite{halder2026local} has investigated energy transfers in statistically steady forced 3D HMHD turbulence, but only at $Pr_m=1$. Their results are qualitatively similar to ours for $Pr_m=1$, but there are quantitative differences that arise from the distinction between decaying and statistically steady turbulence. They do not explore the dependence of these transfers on $Pr_m$. 
\par
A comprehensive understanding of energy transfer mechanisms in 3D HMHD turbulence carries broad implications for astrophysical plasma environments, including the solar wind and magnetospheric plasma, where Hall effects are known to play a significant role. Our work advances understanding of such transfers, particularly by examining the magnetic Prandtl number dependence at both small and large wavenumbers. This is of considerable relevance to both theoretical modeling and observational interpretation of magnetized plasmas.
\par
It is known that Hall-MHD turbulence supports distinct wave modes, including ion-cyclotron waves and whistler waves
[see, e.g., Refs.~\cite{miura2024formation,patel2025uncovering}. 
In future research, we will investigate the energy-transfer processes between the ion-cyclotron and whistler wave modes.

\begin{acknowledgments}
SY and RP thank the Anusandhan National Research Foundation (ANRF), the Science and Engineering Research Board (SERB), and the National Supercomputing Mission (NSM), India, for support,  the Supercomputer Education and Research Centre (IISc), for computational resources, and NIFS (Japan) for hospitality while part of this work was being done. The numerical simulations were performed on an NEC SX-Aurora TSUBASA A412-8 Plasma Simulator at the National Institute of Fusion Science (NIFS), Japan, with the support and under the auspices of the NIFS Collaboration Research program (NIFS23KISS030), FUJITSU FX1000 Wisteria/BDEC-01 Odyssey of the University of Tokyo with partial support from the “Joint Usage/Research Center for Interdisciplinary Large-scale Information Infrastructures” (Project ID:jh230004, jh240004), and FUJITSU/RIKEN FUGAKU with the support of the HPCI System Research Project (Project ID: hp240026) in Japan.
This research was partially supported by JSPS KAKENHI Grant Numbers 24K06893, of Japan, and by the NINS program of Promoting Research by Networking among Institutions (Grant Number 01422301).
This work was supported by the Research Institute for Mathematical Sciences, an International Joint Usage/Research Center located in Kyoto University.
\end{acknowledgments}
\bibliography{aipsamp}

	\end{document}